\documentclass[aps,twocolumn,showpacs,preprintnumbers,amsmath,amssymb,superscriptaddress,nofootinbib]
{revtex4}
\usepackage{graphicx}
\usepackage{dcolumn}
\usepackage{bm}
\usepackage{amsmath}
\usepackage{hyperref}

\def\boldPi{\boldsymbol{\Pi}}
\def\boldxi{\boldsymbol{\xi}}

\def\operatorboost{\widehat{\boldsymbol{\beta}}}
\def\operatorgamma{\widehat{\gamma}}

\def\matrixbeta{\check{\beta}}
\def\matrixalpha{\check{\alpha}}
\def\matrixboldalpha{\check{\boldsymbol{\alpha}}}
\def\matrixboldgamma{\check{\boldsymbol{\gamma}}}
\def\matrixeta{\check{\eta}}

\begin{document}

\title
{Foldy-Wouthuysen transformation for a Dirac-Pauli dyon and the
Thomas-Bargmann-Michel-Telegdi equation}



\author{Tsung-Wei Chen}
\email{twchen@phys.ntu.edu.tw} \affiliation{Department of Physics
and Center for Theoretical Sciences, National Taiwan University,
Taipei 106, Taiwan}

\author{Dah-Wei Chiou}
\email{chiou@gravity.psu.edu}
\affiliation{Department of Physics, Beijing
Normal University, Beijing 100875, China}
\affiliation{Department of Physics
and Center for Theoretical Sciences, National Taiwan University,
Taipei 106, Taiwan}

\begin{abstract}
The classical dynamics for a charged point particle with intrinsic
spin is governed by a relativistic Hamiltonian for the orbital
motion and by the Thomas-Bargmann-Michel-Telegdi equation for the
precession of the spin. It is natural to ask whether the classical
Hamiltonian (with both the orbital and spin parts) is consistent
with that in the relativistic quantum theory for a spin-$1/2$
charged particle, which is described by the Dirac equation. In the
low-energy limit, up to terms of the 7th order in $1/E_g$
($E_g=2mc^2$ and $m$ is the particle mass), we investigate the
Foldy-Wouthuysen (FW) transformation of the Dirac Hamiltonian in
the presence of homogeneous and static electromagnetic fields and
show that it is indeed in agreement with the classical Hamiltonian
with the gyromagnetic ratio being equal to 2. Through
electromagnetic duality, this result can be generalized for a
spin-$1/2$ dyon, which has both electric and magnetic charges and
thus possesses both intrinsic electric and magnetic dipole
moments. Furthermore, the relativistic quantum theory for a
spin-$1/2$ dyon with arbitrary values of the gyromagnetic and
gyroelectric ratios can be described by the Dirac-Pauli equation,
which is the Dirac equation with augmentation for the anomalous
electric and anomalous magnetic dipole moments. The FW
transformation of the Dirac-Pauli Hamiltonian is shown, up to the
7th order again, to be also in accord with the classical
Hamiltonian.
\end{abstract}


\maketitle

\section{Introduction}\label{sec:introduction}
The relativistic quantum theory for a spin-$1/2$ point particle is
described by the Dirac equation~\cite{Dirac28}. The wavefunction
used for the Dirac equation is the Dirac bispinor, which is
composed of two Weyl spinors corresponding to the particle and
antiparticle parts. Rigorously, the Dirac equation is
self-consistent only in the context of quantum field theory, in
which the particle-antiparticle pairs can be created. In the
low-energy limit, if the relevant energy (the particle's energy
interacting with electromagnetic fields) is much smaller than the
Dirac energy gap $E_g=2mc^2$ ($m$ is the particle mass), the
probability of creation of particle-antiparticle pairs is
negligible and the Dirac equation is adequate to describe the
relativistic quantum dynamics of the spin-$1/2$ particle without
taking into account the field-theory interaction to the
antiparticle.

The Foldy-Wouthuysen (FW) transformation is one of
the methods developed to investigate the low-energy limit of the
Dirac equation \cite{Foldy50}.\footnote{It is often said that FW
method gives the nonrelativistic limit of the Dirac equation. The
phrase ``nonrelativistic'' is somewhat misleading as it usually
refers to ``low-speed'' limit. As we will show in this paper, the
FW transformation (if performed to orders high enough) actually
agrees with the relativistic classical dynamics even when the
speed of the particle is large. The appropriate description is to
say that the FW transformation yields ``low-energy'' limit.} In
the FW method, $1/E_g$ is treated as the small parameter; the
Dirac Hamiltonian in the Dirac bispinor representation is block
diagonalized up to a certain order of $1/E_g$ and the remaining
off-diagonal matrices, which correspond to the
particle-antiparticle interactions,  are brought into the next
order of $1/E_g$ and thus neglected. This is achieved by a series
of successive unitary transformations performed on the Dirac
Hamiltonian. Furthermore, a series of successive transformation in
FW method can be reduced into one single transformation by the use
of the L\"{o}wding partitioning method \cite{Lowdin51}.  For a
charged spin-$1/2$ particle subject to a non-explicitly
time-dependent field, an exact FW transformation has been found by
Eriksen \cite{Eriksen58}, and the validity of the transformation
is studied in \cite{Vries68}.

Alternatively, the Dirac Hamiltonian can also be expanded in
powers of Plank constant $\hbar$ \cite{Silenko03}. In this
approach, the small parameter is not the particle's energy
(divided by $E_g$), but its wave length. A diagonalization
procedure based on the expansion in powers of $\hbar$ has been
constructed in \cite{Bliokh05, Goss07}. In this procedure, the
Berry phase correction can also be taken into account.
Furthermore, the semiclassical $\hbar$-expansion enables us to
describe the quantum corrections on the classical expression in
strong fields \cite{Silenko08}.


On the other hand, the classical (non-quantum) dynamics for a
relativistic point particle endowed with charge and intrinsic spin
in static and homogeneous electromagnetic fields is well
understood. The orbital motion is govern by the relativistic
Hamiltonian and the precession of the spin by the
Thomas-Bargmann-Michel-Telegdi (TBMT) equation \cite{BMT59}. The
relativistic Hamiltonian for the orbital motion plus the
Hamiltonian obtained from the TBMT equation (called TBMT
Hamiltonian) is expected to provide a low-energy description of
the relativistic quantum theory. The conjecture that the
low-energy limit of the Dirac Hamiltonian reduces to the classical
orbital Hamiltonian plus the TBMT Hamiltonian has been suggested
but remains to be affirmed.

In order to investigate the
consistency between the low-energy limit of the Dirac equation and
the classical dynamics, we perform a series of FW transformations
and expand the Dirac Hamiltonian up to terms of the 7th order in
$1/E_g$. The electromagnetic fields are assumed to be static and
homogeneous. Taking care of the relation between the kinematic
momentum used in the Dirac Hamiltonian and the boost velocity used
in the TBMT Hamiltonian, we show that the FW transformation of the
Dirac Hamiltonian is in agreement with the classical orbital
Hamiltonian plus the TBMT Hamiltonian for the case of the
gyromagnetic ratio equal to 2. Through electromagnetic duality,
this result can be generalized for a spin-$1/2$ dyon \cite{Shnir},
which has both electric and magnetic charges and thus possesses
both intrinsic electric and magnetic dipole moments (with both
gyromagnetic and gyroelectric ratios equal to 2).

To affirm the consistency to a broader extent, we need to show
that the relativistic quantum theory of a spin-$1/2$ dyon with
arbitrary values of the gyromagnetic and gyroelectric ratios also
reduces to the classical counterparts as a low-energy limit. The
relativistic quantum theory of a spin-$1/2$ dyon with the
inclusion of anomalous magnetic dipole moment (AMM) and anomalous
electric dipole moment (AEM) can be described by the Dirac-Pauli
equation \cite{Silenko08, Pauli41}, which is the Dirac equation
with augmentation for AMM and AEM. The FW transformation is
performed on the Dirac-Pauli Hamiltonian, again up the 7th order
in $1/E_g$, and the result confirms that it remains in agreement
with the classical orbital Hamiltonian plus the TBMT Hamiltonian
for arbitrary values of the gyromagnetic and gyroelectric ratios.

This paper is organized as follows. In Sec.~\ref{sec:dipoles}, we
investigate the tensorial structure of the orbital and intrinsic
dipole moments. In Sec.~\ref{sec:TBMT}, we briefly review the
classical orbital Hamiltonian and the TBMT equation. In
Sec.~\ref{sec:Dirac}, we perform the FW transformation on the
Dirac Hamiltonian for a spin-$1/2$ dyon and show that it agrees
with the TBMT equation for the case of the gyromagnetic and
gyroelectric ratios equal to 2. Later in Sec.~\ref{sec:Dirac
Pauli}, we perform the FW transformation on the Dirac-Pauli
Hamiltonian and show that it again agrees with the TBMT equation
even with the inclusion of AMM and AEM. Finally, the conclusions are
summarized and discussed in Sec.~\ref{sec:conclusions}. Some calculational
details are supplemented in Appendices~\ref{sec:FW transform} and
\ref{sec:derivation}.

\section{Orbital and intrinsic dipole moments}\label{sec:dipoles}
For a general Lorentz transformation from the primed (boosted)
frame to the unprimed (laboratory) frame, the transformation of a
4-vector $\mathrm{k}^\mu=(\mathrm{k}^0,\mathbf{k})$ is given
by~\cite{Jackson}:
\begin{equation}\label{LTboost}
\begin{split}
&\mathrm{k}^{0}=\gamma(\mathrm{k}'^{0}+\boldsymbol{\beta}\cdot\mathbf{k'}),\\
&\mathbf{k}=\mathbf{k'}+\frac{\gamma-1}{\beta^2}(\boldsymbol{\beta}\cdot\mathbf{k'})\boldsymbol{\beta}+\gamma
\mathrm{k}'^0\boldsymbol{\beta},\\
\end{split}
\end{equation}
where $\mathbf{v}=c\boldsymbol{\beta}$ is the boost velocity of the primed frame relative to the unprimed frame, $\gamma$ is the Lorentz factor $\gamma=1/\sqrt{1-\beta^2}$
and $\beta=|\boldsymbol{\beta}|$.

In the primed system, let us consider the case that charge and current densities satisfy the conditions:
\begin{equation}\label{Ncondition}
\int_{V'}d^3x'\rho'=0,
\qquad\int_{V'}d^3x'\mathbf{J}'=0.
\end{equation}
The vanishing of the total charge means that the system is
\emph{neutral}, and the vanishing of the total current is a consequence of the
\emph{static} condition: $\partial\rho'/\partial
t'=-\nabla'\cdot\mathbf{J}'=0$.\footnote{Since the static condition gives $\nabla'\cdot\mathbf{J}'=0$,
it can be shown $J'_i=\nabla'\cdot(x'_i\mathbf{J'})$. Consequently, $\int_{V'}d^3x'J'_i=\int_{V'}d^3x'\,\nabla'\cdot(x'_i\mathbf{J'})
=\int_{\partial V'}d\,\mathbf{a}'\cdot (x'_i\mathbf{J'})=0$
if the current $\mathbf{J}'$ is localized.}
Because the charge density and current density form a 4-vector
$J^\mu=(c\rho,\mathbf{J})$, it can be shown that the same
conditions also hold in the unprimed system:
\begin{equation}
\int_Vd^3x\rho=0,
\qquad
\int_Vd^3x\mathbf{J}=0.
\end{equation}

In the unprimed frame, the magnetic dipole moments $\mathbf{m}$ is
defined as
\begin{equation}
\mathbf{m}=\int_Vd^3x\boldsymbol{\mu}_{m},
\end{equation}
where
\begin{equation}\label{mdmd}
\boldsymbol{\mu}_m=\frac{1}{2c}(\mathbf{x}\times\mathbf{J})
\end{equation}
is the magnetic dipole moment density, and the electric dipole moment
$\mathbf{p}$ is defined as
\begin{equation}
\mathbf{p}=2\int_Vd^3x\boldsymbol{\mu}^c_p,
\end{equation}
where
\begin{equation}\label{cedmd}
\boldsymbol{\mu}^c_p=\frac{1}{2}\mathbf{x}\rho
\end{equation}
is the \emph{canonical} electric dipole moment density (the extra factor of 2 is introduced for later convenience). In the primed
system, the definitions of both dipole moments are the same as those
in the unprimed system. Using Eq.~(\ref{LTboost}), Eq.~(\ref{mdmd}) can
be written as
\begin{equation}
\begin{split}
\boldsymbol{\mu}_m=&\boldsymbol{\mu}'_m+\frac{\gamma-1}{\beta^2}\boldsymbol{\beta}\times(\boldsymbol{\mu}'_m\times\boldsymbol{\beta})\\
&+\frac{1}{2}\gamma(\mathbf{x}'c\rho'-x'^0\mathbf{J}')\times\boldsymbol{\beta},
\end{split}
\end{equation}
where $\boldsymbol{\mu}'_m=\mathbf{x}'\times\mathbf{J}'/2$ is the
magnetic dipole density in the primed system. It is interesting to
note that if we integrate the term
$\mathbf{x}'c\rho'-x'^0\mathbf{J}'$ in the primed system, we
obtain
\begin{equation}
\int_{V'}d^3x'(\mathbf{x}'c\rho'-x'^0\mathbf{J}')=\int_{V'}d^3x'\mathbf{x}'c\rho'
=c\mathbf{p}',
\end{equation}
where the neutral condition [Eq.~(\ref{Ncondition})] has been used. This
suggests that we can define the \emph{tensorial} electric dipole moment as
\begin{equation}
\boldsymbol{\mu}_p=\frac{1}{2c}\left(\mathbf{x}J^0-x^0\mathbf{J}\right)
\end{equation}
so that the dipole moment can be defined as a second rank antisymmetric tensor:
\begin{equation}\label{DipoleTensor}
M^{\mu\nu}=\frac{1}{2c}(x^{\mu}J^{\nu}-x^{\nu}J^{\mu}).
\end{equation}
The canonical and tensorial dipole moments densities yield the same (integrated) dipole moments, because the neutral condition ensures that the integration of the second
term $x^0\mathbf{J}$ vanishes in the unprimed system. The components
of the second rank tensor $M^{\mu\nu}$ are
\begin{equation}
\begin{split}
&M^{0i}=\frac{1}{2c}(x^0J^i-x^iJ^0)=-\mu_p^i,\\
&M^{ij}=\frac{1}{2c}(x^iJ^j-x^jJ^i)=\epsilon_{ijk}\mu_m^k.
\end{split}
\end{equation}
Consequently, the Lorentz transformation between
$(\boldsymbol{\mu}_p,\boldsymbol{\mu}_m)$ and
$(\boldsymbol{\mu}'_p,\boldsymbol{\mu}'_m)$ is of the form
\begin{equation}\label{LTdipole}
\begin{split}
&\boldsymbol{\mu}_p=\gamma(\boldsymbol{\mu}'_p+\boldsymbol{\beta}\times\boldsymbol{\mu}'_m)-\frac{\gamma^2}{\gamma+1}\boldsymbol{\beta}(\boldsymbol{\beta}\cdot\boldsymbol{\mu}'_p),\\
&\boldsymbol{\mu}_m=\gamma(\boldsymbol{\mu}'_m-\boldsymbol{\beta}\times\boldsymbol{\mu}'_p)-\frac{\gamma^2}{\gamma+1}\boldsymbol{\beta}(\boldsymbol{\beta}\cdot\boldsymbol{\mu}'_m).
\end{split}
\end{equation}
The transformation [Eq.~(\ref{LTdipole})] is exactly the same as
that for the electric and magnetic fields if we take the
replacement rules: $\mathbf{E}\leftrightarrow
\boldsymbol{\mu}_p$\,
$\mathbf{B}\leftrightarrow-\boldsymbol{\mu}_m$.

The corresponding transformation for the (integrated) dipole moments is given
by
\begin{equation}\label{LTdipole2}
\begin{split}
&\frac{\mathbf{p}}{2}=\gamma^2\left[\frac{\mathbf{p}'}{2}+\boldsymbol{\beta}\times\mathbf{m}'-\frac{\gamma}{\gamma+1}\boldsymbol{\beta}(\boldsymbol{\beta}\cdot\frac{\mathbf{p}'}{2})\right],\\
&\mathbf{m}=\gamma^2\left[\mathbf{m}'-\boldsymbol{\beta}\times\frac{\mathbf{p}'}{2}-\frac{\gamma}{\gamma+1}\boldsymbol{\beta}(\boldsymbol{\beta}\cdot\mathbf{m}')\right],
\end{split}
\end{equation}
where $d^3x=\gamma d^3x'$ is used. Since the extra factor $\gamma$
arises in the right hand side of Eq.~(\ref{LTdipole2}) due to the
spatial integral, $\mathbf{p}/2$ and $\mathbf{m}$ do not transform
covariantly and thus do not form a second rank tensor unlike
$\boldsymbol{\mu}_p$ and $\boldsymbol{\mu}_m$.

In the case with only proper electric dipole moment and no proper
magnetic dipole moment (i.e. $\mathbf{p}'\neq0$ and
$\mathbf{m}'=0$), when boosted, the electric dipole will result in
a magnetic dipole moment
$\mathbf{m}=-\gamma^2(\boldsymbol{\beta}\times\mathbf{p}'/2)$ in
the unprimed system. This can be understood as follows: If we
think $\boldsymbol{p}'$ as two endpoints separated by a short
distance and charged with $+q$ and $-q$, in the unprimed system,
the positive and negative charges acquire a velocity and give rise
to currents in opposite directions, thus resulting in a magnetic
dipole moment. In an inhomogeneous magnetic field, a moving object
with only proper electric dipole moment can feel the magnetic
force $\mathbf{F}=(\mathbf{m}\cdot\nabla)\mathbf{B}$.

On the other hand, in the case with only proper magnetic dipole
moment and no proper electric dipole moment (i.e.
$\mathbf{m}'\neq0$ and $\mathbf{p}'=0$), when boosted, the
magnetic dipole will result in an electric dipole moment
$\mathbf{p}=2\gamma^2(\boldsymbol{\beta}\times\mathbf{m}')$ in the
unprimed system. This is due to the fact that in the unprimed
system the charge density $\rho$ arises through the Lorentz
transformation even if the charge density is zero in the primed
system ($\rho'=0$). The charge density in the primed system
$c\rho=\gamma\boldsymbol{\beta}\cdot\mathbf{J}'$ is positive
(negative) when the current is parallel (anti-parallel) to the
boost velocity; therefore, as a magnetic dipole can be thought as
a small current loop, the small current loop in the primed system
gives rise to opposite charges separated by a short distance in
the unprimed system, thus resulting in an electric dipole moment.
In an inhomogeneous electric field, a moving object with only
proper magnetic dipole moment can feel the electric force
$\mathbf{F}=(\mathbf{p}\cdot\nabla)\mathbf{E}$.

The dipole moments considered above are \emph{orbital} in the sense that they are sourced by the orbital distribution of $J^\mu(x)$. On the other hand, a point particle can give rise to an \emph{intrinsic} dipole moment if it is charged and endowed with intrinsic spin. The fact that $\boldsymbol{\mu}_p$ and $\boldsymbol{\mu}_m$ form an antisymmetric tensor $M^{\mu\nu}$ suggests that the intrinsic spin $\mathbf{s}$ can be generalized to a second-rank antisymmetric tensor $S^{\mu\nu}$, which gives the intrinsic dipole moments as
\begin{equation}\label{M and S}
M^{\mu\nu}=\frac{g_ee}{2mc}\,S^{\mu\nu},
\end{equation}
where $e$ is the electric charge of the particle, $m$ the mass and $g_e$ the \emph{gyromagnetic ratio}. The spin has only three independent components; thus $S^{\mu\nu}$ is dual to an axial 4-vector $S^\alpha=(S^0,\mathbf{S})$ via
\begin{equation}
S^{\mu\nu}=\frac{1}{c}\,\epsilon^{\mu\nu\alpha\beta}U_\alpha S_\beta
\end{equation}
and conversely
\begin{equation}
S^\alpha=\frac{1}{2c}\,\epsilon^{\alpha\beta\gamma\delta}U_\beta S_{\gamma\delta},
\end{equation}
where $U^\alpha$ is the particle's 4-velocity. The 4-vector $S^\alpha$ reduces to the spin $\mathbf{s}$ in the particle's rest frame; i.e., $S'^\alpha=(S'^0,\mathbf{S}')=(0,\mathbf{s})$. The vanishing of the time-component in the particle's rest frame is imposed by the covariant constraint:
\begin{equation}
U_\alpha S^\alpha=0.
\end{equation}
In the particle's rest frame, $U^{\prime\alpha}=(c,0,0,0)$ and Eq.~(\ref{M and S}) yields
\begin{equation}
\boldsymbol{\mu}'_m=\frac{g_e e}{2mc}\,\mathbf{s},
\qquad
\boldsymbol{\mu}'_p=0.
\end{equation}
Therefore, the intrinsic spin gives only the proper intrinsic
magnetic dipole and no proper intrinsic electric dipole. In order
to have both proper intrinsic magnetic and electric dipoles, we
consider a \emph{dyon} particle~\cite{Shnir}, which possesses both
electric charge $e$ and magnetic charge $\tilde{e}$, and
Eq.~(\ref{M and S}) is generalized as
\begin{equation}\label{M and S S}
M^{\mu\nu}=M^{\mu\nu}_e+M^{\mu\nu}_{\tilde{e}}=\frac{g_ee}{2mc}\,S^{\mu\nu}
+\frac{g_{\tilde{e}}\tilde{e}}{2mc}\,\tilde{S}^{\mu\nu},
\end{equation}
where $g_{\tilde{e}}$ is the \emph{gyroelectirc ratio} and
\begin{equation}
\tilde{S}^{\mu\nu}:=\frac{1}{2}\,\epsilon^{\mu\nu\alpha\beta}S_{\alpha_\beta}
\end{equation}
is the dual of $S^{\mu\nu}$. In the rest frame, Eq.~(\ref{M and S S}) yields both magnetic and electric dipoles:
\begin{equation}
\boldsymbol{\mu}^{\prime e}_m=\frac{g_e e}{2mc}\,\mathbf{s},
\qquad
\boldsymbol{\mu}^{\prime \tilde{e}}_p=-\frac{g_{\tilde{e}} \tilde{e}}{2mc}\,\mathbf{s}.
\end{equation}

\section{The Thomas-Bargmann-Michel-Telegdi equation}\label{sec:TBMT}
Consider a relativistic point particle endowed with electric charge and intrinsic spin subject to static and homogeneous electromagnetic fields. The orbital motion of the particle is described by
\begin{equation}\label{covariant eom 1}
\frac{dU^\alpha}{d\tau}=\frac{e}{mc}F^{\alpha\beta}U_\beta
\end{equation}
and the precession of the spin is govern by the TBMT equation \cite{BMT59}:
\begin{equation}\label{covariant eom 2}
\frac{dS^\alpha}{d\tau}=\frac{e}{mc}
\left[
\frac{g_e}{2}\,F^{\alpha\beta}S_\beta
+\frac{1}{c^2}\left(\frac{g_e}{2}-1\right)U^\alpha
\left(S_\lambda F^{\lambda\mu} U_\mu\right)
\right].
\end{equation}

Equation (\ref{covariant eom 1}) in the covariant form can be shown to be equivalent to the Hamilton's equations :
\begin{equation}
\begin{split}
\frac{d\mathbf{x}}{dt}&=\{\mathbf{x},H_\mathrm{oribt}\},\\
\frac{d\mathbf{p}}{dt}&=\{\mathbf{p},H_\mathrm{orbit}\}
\end{split}
\end{equation}
in the unprimed frame, where $\mathbf{p}$ is the conjugate momentum to $\mathbf{x}$ and the Hamiltonian $H_\mathrm{orbit}$ governing the orbital motion is given by
\begin{equation}\label{H orbit}
H_\mathrm{orbit}(\mathbf{x},\mathbf{p})
=\sqrt{\left(c\,\mathbf{p}-e\mathbf{A}(\mathbf{x})\right)^2+m^2c^4}\,
+e\,\phi(\mathbf{x})
\end{equation}
with $A^\alpha=(\phi,\mathbf{A})$ being the 4-vector potential for the electromagnetic field $F^{\mu\nu}$. (See Sec.~12.1 in \cite{Jackson} for more details.)

On the other hand, Eq.~(\ref{covariant eom 2}) leads to
\begin{equation}\label{Thomas}
\frac{d\mathbf{s}}{dt}=\frac{e}{mc}\,\mathbf{s}\times\mathbf{F}(\mathbf{x})
\end{equation}
with
\begin{equation}\label{ThomasF}
\begin{split}
\mathbf{F}&=\left(\frac{g_e}{2}-1+\frac{1}{\gamma}\right)\mathbf{B}-\left(\frac{g_e}{2}-1\right)\frac{\gamma}{\gamma+1}(\boldsymbol{\beta}\cdot\mathbf{B})\boldsymbol{\beta}\\
&~~-\left(\frac{g_e}{2}-\frac{\gamma}{\gamma+1}\right)\boldsymbol{\beta}\times\mathbf{E},
\end{split}
\end{equation}
which gives the spin precession with respect to the time of the unprimed frame. (See Sec.~11.11 in \cite{Jackson} for more details.) Because $\{s_i,s_j\}=\epsilon_{ijk}s_k$, Eq.~(\ref{Thomas}) can be recast as the Hamilton's equation:
\begin{equation}
\frac{d\mathbf{s}}{dt}=\{\mathbf{s},H_\mathrm{spin}\}
\end{equation}
with
\begin{equation}\label{H dipole}
H_\mathrm{spin}(\mathbf{x},\mathbf{s})=-\frac{e}{mc}\,\mathbf{s}\cdot \mathbf{F}(\mathbf{x})
\end{equation}
called the TBMT Hamiltonian, which governs the precession of the
electric dipole subject to a static and homogeneous field.

In the low-speed limit ($\beta\ll 1$), we have $\gamma\approx1$
and Eq.~(\ref{H dipole}) gives
\begin{eqnarray}\label{low speed H dipole}
H_\mathrm{spin}
&\approx&-\frac{e}{2mc}\,\mathbf{s}\cdot
\bigg[
g_e\mathbf{B}
-\left(\frac{g_e}{2}-1\right)(\boldsymbol{\beta}\cdot\mathbf{B})\boldsymbol{\beta}\nonumber\\
&&\qquad\qquad-(g_e-1)\boldsymbol{\beta}\times\mathbf{E}
\bigg].
\end{eqnarray}
The first term in Eq.~(\ref{low speed H dipole}) is the
interaction energy of the magnetic moment
$\boldsymbol{\mu}^{\prime e}_m$ in the magnetic field, which
accounts for the anomalous Zeeman effect. The second term
corresponds to the change rate of the longitudinal polarization,
which vanishes in the case of $g_e=2$. The third term is the
spin-orbit interaction (the interaction of the boosted electric
dipole
$\boldsymbol{\mu}^{e}_p\approx\boldsymbol{\beta}\times\boldsymbol{\mu}^{\prime
e}_m$ coupled to the electric field) plus the correction for the
Thomas precession.

By treating $\mathbf{x}$, $\mathbf{p}$ and $\mathbf{s}$ as
independent phase space variables, the total Hamiltonian is given
by\footnote{Note that, in order to add $H_\mathrm{orbit}$ and
$H_\mathrm{spin}$ together, we have to consider
$d\mathbf{s}/dt\equiv d\mathbf{S}'/dt$ in Eq.~(\ref{Thomas}),
instead of $d\mathbf{S}/dt$, $d\mathbf{S}/d\tau$ or
$d\mathbf{s}/d\tau$. This is because $s_i$ are degrees of freedom
independent of $\mathbf{x}$ and $\mathbf{p}$, but $S_i$ are not.
Furthermore, to be consistent with the orbital motion, the
precession is cast with respect to $t$, instead of the proper time
$\tau$ of the moving particle.}
\begin{equation}\label{H total}
H(\mathbf{x},\mathbf{p},\mathbf{s})=H_\mathrm{orbit}(\mathbf{x},\mathbf{p})
+H_\mathrm{spin}(\mathbf{x},\mathbf{s}).
\end{equation}
If the particle has both electric charge $e$ and magnetic charge $\tilde{e}$ (i.e. the particle is a \emph{dyon}), Eq.~(\ref{H orbit}) and Eq.~(\ref{H dipole}) are modified with the inclusion of the dual counterparts; i.e.
\begin{eqnarray}\label{H orbit dyon}
H_\mathrm{orbit}(\mathbf{x},\mathbf{p})
&=&\sqrt{\left(c\,\mathbf{p}-e\mathbf{A}(\mathbf{x})-\tilde{e}\tilde{\mathbf{A}}(\mathbf{x})
\right)^2+m^2c^4}\nonumber\\
&&+\,e\phi(\mathbf{x})+\tilde{e}\tilde{\phi}(\mathbf{x})
\end{eqnarray}
and
\begin{equation}\label{H dipole dyon}
H_\mathrm{spin}(\mathbf{x},\mathbf{s})=-\frac{e}{mc}\,\mathbf{s}\cdot \mathbf{F}(\mathbf{x})
-\frac{\tilde{e}}{mc}\,\mathbf{s}\cdot \tilde{\mathbf{F}}(\mathbf{x})
\end{equation}
with
\begin{equation}\label{ThomasF dual}
\begin{split}
\tilde{\mathbf{F}}&=\left(\frac{g_{\tilde{e}}}{2}-1+\frac{1}{\gamma}\right)\tilde{\mathbf{B}}
-\left(\frac{g_{\tilde{e}}}{2}-1\right)
\frac{\gamma}{\gamma+1}(\boldsymbol{\beta}\cdot\tilde{\mathbf{B}})\boldsymbol{\beta}\\
&~~-\left(\frac{g_{\tilde{e}}}{2}-\frac{\gamma}{\gamma+1}\right)
\boldsymbol{\beta}\times\tilde{\mathbf{E}},
\end{split}
\end{equation}
where $\tilde{A}=(\tilde{\phi},\tilde{\mathbf{A}})$ is the dual 4-vector potential which gives $\tilde{F}^{\mu\nu}=\partial^\mu \tilde{A}^\nu-\partial^\nu \tilde{A}^\mu$ and $\tilde{F}^{\mu\nu}:=1/2\,\epsilon^{\mu\nu\alpha\beta}F_{\alpha\beta}$ is the dual field strength (i.e. $\tilde{\mathbf{B}}=-\mathbf{E}$ and $\tilde{\mathbf{E}}=\mathbf{B}$).

Equation~(\ref{covariant eom 1}) and the TBMT equation given in
Eq.~(\ref{covariant eom 2}) are derived as the requirement of
covariant is considered. They are classical (non-quantum)
equations and we wonder whether the Hamiltonian given in
Eq.~(\ref{H total}) is consistent with that in the relativistic
quantum theory for a charged point particle with intrinsic spin.

The relativistic quantum theory of a spin-$1/2$ particle is
described by the Dirac equation. The Dirac bispinor however has
both the particle and antiparticle components, which are entangled
by the Dirac equation.

In order to compare with the TBMT equation, we consider the
low-energy limit in which the relevant energy is much smaller than
the Dirac energy gap $E_g$ and the FW transformation is used to
block-diagonalize the Dirac Hamiltonian. In
Sec.~{\ref{sec:Dirac}}, we will show that the FW transformation of
the Dirac Hamiltonian indeed agrees perfectly with the TBMT
equation up to the 7th order of $1/E_g$ with the intrinsic spin
given by $\mathbf{s}=\hbar\,\boldsymbol{\sigma}/2$ ($\sigma_i$ are
the Pauli matrices) and the gyromagnetic ratio given by $g_e=2$.
This can be easily generalized for a Dirac dyon by adding the
magnetic charge (and we will have $g_e=g_{\tilde{e}}=2$).

As the Dirac equation always yields $g_e=2$, we will not see the
second term in Eq.~(\ref{ThomasF}), which accounts for change of
the longitudinal polarization. In order to see that the quantum
theory is in accord with the TBMT equation even for the case of
$g_e\neq2$, we study the Dirac-Pauli equation in
Sec.~\ref{sec:Dirac Pauli} with the inclusion of anomalous dipole
moments. The results again affirms the consistency between the FW
transformation of the Dirac-Pauli Hamiltonian and the TBMT
equation up to the 7th order of $1/E_g$.

\section{Foldy-Wouthuysen transformation for the Dirac Hamiltonian}\label{sec:Dirac}
The relativistic quantum theory of a Dirac particle is described by the Dirac equation
\begin{equation}
i\hbar\frac{\partial}{\partial t}|\psi\rangle=H|\psi\rangle,
\end{equation}
where the Dirac bispinor $|\psi\rangle=(\chi,\varphi)^T$ is composed of two 2-component Weyl spinors $\chi$ and $\varphi$ corresponding to the particle and antiparticle parts. The Dirac Hamiltonian is given by
\begin{equation}\label{H}
H=mc^2\matrixbeta+c\,\matrixboldalpha\cdot\boldsymbol{\Pi}+V,
\end{equation}
where the $4\times4$ matrices
$\matrixbeta=\sigma_z\otimes\mathbf{1}$ and
$\matrixalpha_i=\sigma_x\otimes\sigma_i$ are given in the
Pauli-Dirac representation~\cite{Dirac28} and satisfy\footnote{To
avoid confusion with the boost velocity $\beta$, we use the
checked notation $\matrixbeta$ to denote the $4\times4$ matrix.
With the same style, the notations $\matrixalpha$,
$\matrixboldgamma$ [defined in Eq.~(\ref{matrices beta and
gamma})] and $\matrixeta$ [defined in Eq.~(\ref{matrix eta})] are
checked as well.}
\begin{equation}
\begin{split}
&\{\matrixbeta,\matrixalpha_i\}=0,\\
&\{\matrixalpha_i,\matrixalpha_j\}=2\delta_{ij},\\
&\matrixalpha_i^2=\matrixbeta^2=\mathbf{1}.
\end{split}
\end{equation}
The mass of the particle is $m$ and $\boldsymbol{\Pi}$ is the kinetic momentum.

In order to have the proper intrinsic electric dipole moment and the proper intrinsic magnetic dipole moment at the same time, we consider a Dirac dyon (i.e. a Dirac particle with both electric charge $e$ and magnetic charge $\tilde{e}$). For a dyon \cite{Shnir}, the kinetic momentum is given by
\begin{equation}
\boldsymbol{\Pi}=\mathbf{p}-\frac{e}{c}\mathbf{A}-\frac{\tilde{e}}{c}\widetilde{\mathbf{A}},
\end{equation}
and the scalar potential $V$ is composed of electric and magnetic monopole potentials:
\begin{equation}
V=e\phi+\tilde{e}\widetilde{\phi}\,.
\end{equation}

In the following, we will first perform the successive FW transformations of the Dirac Hamiltonian up to the 7th order of $1/E_g$ in Sec.~\ref{sec:FW}, and later show that the results agree with the TBMT equation in Sec.~\ref{sec:relation to TBMT}.

\subsection{Foldy-Wouthuysen transformation}\label{sec:FW}
In order to perform the FW transformation~\cite{Foldy50}, we have to rewrite the Dirac Hamiltonian to the form:
\begin{equation}\label{H_D2}
H=E_g\frac{\matrixbeta}{2}+\Omega_o+\Omega_E,
\end{equation}
where $E_g=2mc^2$ is the Dirac energy gap, and the odd matrix
$\Omega_o$ and the even matrix $\Omega_E$ are defined as
\begin{equation}\label{OE}
\{\matrixbeta,\Omega_o\}=0,\qquad[\matrixbeta,\Omega_E]=0.
\end{equation}
In the case of Eq.~(\ref{H}), we have
\begin{equation}\label{OE1}
\Omega_o=c\,\matrixboldalpha\cdot\boldsymbol{\Pi},
\qquad\Omega_E=V.
\end{equation}

The resulting effective hamiltonian $H_\mathrm{FW}$ can be obtained by
the successive unitary transformations which partitioning off the
odd matrices to a higher order. In general, we can use the FW
matrix $U_\mathrm{FW}$ as a single transformation, and expand the
exponent of the matrix in powers of $1/E_g$; this is the
well-known L\"{o}wdin partitioning method~\cite{Lowdin51}. It
can be shown that the FW transformation of Eq.~(\ref{H_D2}),
namely the transformed Hamiltonian denoted as
$H_\mathrm{FW}=U_\mathrm{FW}HU_\mathrm{FW}^{-1}$, up to terms of the 7th order in
$1/E_g$ can be written as (see Appendix~\ref{sec:FW transform})
\begin{equation}\label{HFW}
H_\mathrm{FW}=\frac{\matrixbeta
E_g}{2}+\Omega_E+\sum_{\ell=1}^{6}H^{(\ell)}_\mathrm{FW}+o(1/E_g^7),
\end{equation}
where the first four terms $H_\mathrm{FW}^{(\ell)},~\ell=1,2,3,4$ are
given by
\begin{subequations}\label{HFW_list}
\begin{align}
H^{(1)}_\mathrm{FW}&=\frac{\matrixbeta\Omega_o^2}{E_g}\label{HFW_list1},\\
H^{(2)}_\mathrm{FW}&=\frac{1}{E_g^2}\left(\frac{\mathcal{W}}{2}\right)\label{HFW_list2},\\
H^{(3)}_\mathrm{FW}&=\frac{1}{E_g^3}\left\{-\matrixbeta\Omega_o^4+\matrixbeta
\left(\matrixbeta\mathcal{D}\right)^2\right\}\label{HFW_list3},\\
H^{(4)}_\mathrm{FW}&=\frac{1}{E_g^4}\left(\frac{1}{24}[[\Omega_o,\mathcal{W}]
\Omega_o]-\frac{4}{3}[\mathcal{D},\Omega_o^3]\right)\label{HFW_list4},
\end{align}
\end{subequations}
and the operators $\mathcal{W}$ and $\mathcal{D}$ are defined as
\begin{subequations}\label{DandW}
\begin{align}
&\mathcal{D}=[\Omega_o,\Omega_E],\\
&\mathcal{W}=[\mathcal{D},\Omega_o].
\end{align}
\end{subequations}

By using Eq.~(\ref{OE1}), it can be shown the three Hamiltonians
$H_\mathrm{FW}^{(\ell=1,2,3)}$ are in agreement with the previous
results~\cite{Foldy50, Froh93}. The term of the 5th order is given
by
\begin{equation}\label{HFW_list5}
\begin{split}
E_g^5H^{(5)}_\mathrm{FW}&=\frac{1}{144}[(\matrixbeta\Omega_o)_{(5)},\Omega_o]+\frac{1}{2}\mathop{\sum_{\ell,m=1}^{3}}\limits_{(\ell+m=4)}[\matrixbeta\mathcal{O}^{(\ell)},\mathcal{O}^{(m)}]\\
&~~+\frac{1}{2}\mathop{\sum_{\ell,m=1}^{2}\sum_{n=0}^{1}}\limits_{(\ell+m+n=3)}[\matrixbeta\mathcal{O}^{(\ell)},[\matrixbeta\mathcal{O}^{(m)},h^{(n)}]],
\end{split}
\end{equation}
where the subscript $(5)$ in the commutator
$[(\matrixbeta\Omega_o)_{(5)},\Omega_o]$ indicates that the
commutation of $\matrixbeta\Omega_o$ with $\Omega_o$ is performed
successively by five times; i.e.,
$[(\matrixbeta\Omega_o)_{(5)},\Omega_o] =
[\matrixbeta\Omega_o,[\matrixbeta\Omega_o,[\matrixbeta\Omega_o,[\matrixbeta\Omega_o[\matrixbeta\Omega_o,\Omega_o]]]]]$.
The term of the 6th order is
\begin{equation}\label{HFW_list6}
\begin{split}
E_g^6H^{(6)}_\mathrm{FW}&=\frac{1}{720}[(\matrixbeta\Omega_o)_{(6)},\Omega_E]+\frac{1}{2}\mathop{\sum_{\ell,m=1}^{4}}\limits_{(\ell+m=5)}[\matrixbeta\mathcal{O}^{(\ell)},\mathcal{O}^{(m)}]\\
&~~+\frac{1}{2}\mathop{\sum_{\ell,m=1}^{3}\sum_{n=0}^{2}}\limits_{(\ell+m+n=4)}[\matrixbeta\mathcal{O}^{(\ell)},[\matrixbeta\mathcal{O}^{(m)},h^{(n)}]],
\end{split}
\end{equation}
where the odd matrices $\mathcal{O}^{(\ell)}$ for $\ell=1,2,3,4$ are
given by
\begin{equation}\label{listO}
\begin{split}
&\mathcal{O}^{(1)}=\matrixbeta\mathcal{D},\quad\mathcal{O}^{(2)}=-\frac{4}{3}\Omega_o^3,\\
&\mathcal{O}^{(3)}=\frac{1}{6}\matrixbeta[\Omega_o,\mathcal{W}],\quad
\mathcal{O}^{(4)}=\frac{8}{15}\Omega_o^5,
\end{split}
\end{equation}
and the even matrices $h^{(n)}$ for $n=0,1,2$ are
\begin{equation}
\begin{split}
&h^{(0)}=\Omega_E,\quad h^{(1)}=\matrixbeta\Omega_o^2,\\
&h^{(2)}=\frac{\mathcal{W}}{2}.
\end{split}
\end{equation}
To obtain the FW transformed Hamiltonian $H_\mathrm{FW}^{(\ell)}$ up
to the 7th order of $1/E_g$, we need only three successive transformations $U_\mathrm{FW}=\exp(S_3)\exp(S_2)\exp(S_1)$ (see Appendix~\ref{sec:FW transform}), and it can be shown that $S_1$, $S_2$ and $S_3$ are all
anti-hermitian matrices.


To simplify the calculation, some restrictions and
assumptions are made. The electromagnetic
field is assumed to be static, as has been used in obtaining Eq.~(\ref{HFW}). In order to demonstrate the equivalence clearly
between the TBMT Hamiltonian and $H_\mathrm{FW}$, we further assume that the
external fields $\mathbf{E}$ and $\mathbf{B}$ are homogeneous, and
thus the field gradient vanishes. Furthermore, the terms
proportional to products of field strengths, such as $E_iE_j$,
$E_iB_j$ and $B_iB_j$, are all neglected as a good approximation for weak fields.


We now evaluate each term of $H^{(\ell)}_\mathrm{FW}$. The kinetic term
$\Omega_o^2/E_g$ in Eq.~(\ref{HFW_list1}) can be written as
\begin{equation}\label{Omega2}
\begin{split}
\frac{\Omega_o^2}{E_g}&=\frac{(c\matrixboldalpha\cdot\boldsymbol{\Pi})^2}{2mc^2}\\
&=\frac{1}{2m}\left\{|\boldsymbol{\Pi}|^2+i\boldsymbol{\Sigma}\cdot\left(\boldsymbol{\Pi}\times\boldsymbol{\Pi}\right)\right\},
\end{split}
\end{equation}
where $[\matrixalpha_i,\matrixalpha_j]=2i\epsilon_{ijk}\Sigma_k$
is used. By using the definition of magnetic field
$\mathbf{B}=\nabla\times\mathbf{A}$ and dual magnetic field
$\widetilde{\mathbf{B}}=\nabla\times\widetilde{\mathbf{A}}$, we
have
$c\,\boldsymbol{\Pi}\times\boldsymbol{\Pi}=i\hbar(e\mathbf{B}+\tilde{e}\widetilde{\mathbf{B}})$.
By applying the duality $\widetilde{\mathbf{B}}=-\mathbf{E}$ in
Eq.~(\ref{Omega2}), Eq.~(\ref{HFW_list1}) then gives
\begin{equation}\label{HFW1}
H^{(1)}_\mathrm{FW}=\frac{\matrixbeta|\boldsymbol{\Pi}|^2}{2m}
-\matrixbeta\left(\frac{e\hbar}{2mc}\boldsymbol{\Sigma}\right)\cdot\mathbf{B}
-\matrixbeta\left(-\frac{\tilde{e}\hbar}{2mc}\boldsymbol{\Sigma}\right)\cdot\mathbf{E}.
\end{equation}
The second term of Eq.~(\ref{HFW1}) is the Zeeman Hamiltonian for an electron (with $e=-|e|$) \cite{Sakurai}, and the third term is its duality. It is
interesting to note that $\tilde{e}\hbar\boldsymbol{\Sigma}/2mc$
plays the role of electric dipole moment because it couples to the
electric field. In this sense, we can define the (proper) intrinsic
electric dipole moment ($\boldsymbol{\mu}_p^{\prime\tilde{e}}$) and (proper)
intrinsic magnetic dipole moment ($\boldsymbol{\mu}_m^{\prime e}$) as
\begin{subequations}\label{Intrinsicdipoles}
\begin{align}
\boldsymbol{\mu}_m^{\prime e}&=\frac{e\hbar}{2mc}\boldsymbol{\Sigma}\label{IMDM},\\
\boldsymbol{\mu}_p^{\prime\tilde{e}}&=
-\frac{\tilde{e}\hbar}{2mc}\boldsymbol{\Sigma}\label{IEDM}.
\end{align}
\end{subequations}
Equation~(\ref{IMDM}) implies that the dyon's intrinsic
gyromagnetic ratio $g_e=2$ for the Dirac Hamiltonian. We also find the
same gyroelectric ratio $g_{\tilde{e}}=2$ for the dyon's intrinsic
electric dipole moment.

We now focus on the 2nd-order Hamiltonian $H_\mathrm{FW}^{(2)}$ in
which the spin-orbit coupled term is included. It can be shown
that $\mathcal{W}$ is given by
$\mathcal{W}=-2c^2\hbar\boldsymbol{\Sigma}\cdot
((e\mathbf{E}+\tilde{e}\widetilde{\mathbf{E}})\times\boldsymbol{\Pi})$
and Eq.~(\ref{HFW_list2}) can be written as
\begin{equation}\label{HFW2}
H_\mathrm{FW}^{(2)}
=-\frac{1}{2}\mathbf{E}\cdot
\left(\frac{\boldsymbol{\Pi}}{mc}\times\boldsymbol{\mu}_m^{\prime e}\right)
-\frac{1}{2}\mathbf{B}\cdot
\left(-\frac{\boldsymbol{\Pi}}{mc}\times\boldsymbol{\mu}_p^{\prime \tilde{e}}\right),
\end{equation}
where the duality $\tilde{\mathbf{E}}=\mathbf{B}$
is used. The first term of Eq.~(\ref{HFW2}) is the spin-orbit
interaction for an electron~\cite{Sakurai}. On the other hand, we
neglect the terms proportional to products of field strengths in
evaluating the two terms in Eq.~(\ref{HFW_list3}), and thus we can
obtain
\begin{equation}\label{HFW3}
\begin{split}
H_\mathrm{FW}^{(3)}
&\approx-\frac{\matrixbeta|\boldsymbol{\Pi}|^4}{8m^3c^2}
+\frac{1}{2}\matrixbeta\left(\frac{|\boldsymbol{\Pi}|}{mc}\right)^2
\left(\boldsymbol{\mu}_m^{\prime e}\cdot\mathbf{B}\right)\\
&~~~~+\frac{1}{2}\matrixbeta
\left(\frac{|\boldsymbol{\Pi}|}{mc}\right)^2
(\boldsymbol{\mu}_p^{\prime \tilde{e}}\cdot\mathbf{E}),
\end{split}
\end{equation}
where the assumption of homogeneous electromagnetic fields is
used, and thus the operator $|\boldsymbol{\Pi}|^2$ commutes with the
magnetic field $\mathbf{B}$. If the magnetic charge $\tilde{e}$
vanishes, the first term of Eq.~(\ref{HFW3}) is the relativistic
mass correction that contributes to the spectrum of fine
structure~\cite{Sakurai}.

The second term of Eq.~(\ref{HFW3}) is the relativistic correction
to the Zeeman Hamiltonian appearing in $H_\mathrm{FW}^{(1)}$
[Eq.~(\ref{HFW1})]. For the 4th order $H^{(4)}_\mathrm{FW}$, it
can be shown that
\begin{equation}\label{HFW_4_1}
\begin{split}
&\frac{1}{24}[[\Omega_o,\mathcal{W}],\Omega_o]-\frac{4}{3}[[\Omega_o,\Omega_e],\Omega_o^3]\\
&=-\frac{11}{8}(\Omega_o^2\mathcal{W}+\mathcal{W}\Omega_o^2)
-\frac{5}{4}\Omega_o\mathcal{W}\Omega_o.
\end{split}
\end{equation}
If we neglect all terms proportional to products of
electromagnetic fields, one can show that (see Appendix~\ref{sec:derivation}):
\begin{subequations}\label{HFW_4_2}
\begin{align}
&(\Omega_o^2\mathcal{W}+\mathcal{W}\Omega_o^2)\approx 2c^2|\boldsymbol{\Pi}|^2\mathcal{W},\label{HFW_4_2(1)}\\
&\Omega_o\mathcal{W}\Omega_o\approx-c^2|\boldsymbol{\Pi}|^2\mathcal{W}\label{HFW_4_2(2)}.
\end{align}
\end{subequations}
Note that there is a minus sign in Eq.~(\ref{HFW_4_2(2)}). The
4th-order term Eq.~(\ref{HFW_list4}) with substitution of Eqs.~(\ref{HFW_4_1}) and (\ref{HFW_4_2}) can be written as
\begin{equation}\label{HFW4}
H^{(4)}_\mathrm{FW}\approx\frac{c^2}{E_g^4}
\left(-\frac{3}{2}\right)|\boldsymbol{\Pi}|^2\mathcal{W}
=-\frac{3}{4}\left(\frac{|\boldsymbol{\Pi}|}{mc}\right)^2H^{(2)}_\mathrm{FW}.
\end{equation}
It is interesting to note that the 4th order Hamiltonian
$H_\mathrm{FW}^{(4)}$ is in relation to the 2nd order Hamiltonian
$H_\mathrm{FW}^{(2)}$ by a relativistic correction
$-3(|\boldPi|/mc)^2/4$. For those terms in the 5th order, it can be
sown that each term corresponding to Eq.~(\ref{HFW_list5}) is
given by
\begin{equation}\label{HFW_5_1}
\begin{split}
&\quad[(\matrixbeta\Omega_o)_{(5)},\Omega_o]=32\matrixbeta\Omega_o^6,\\
&\mathop{\sum_{\ell,m=1}^{3}}\limits_{(\ell+m=4)}
[\matrixbeta\mathcal{O}^{(\ell)},\mathcal{O}^{(m)}]
=-\frac{1}{3}\matrixbeta\{\mathcal{D},[\Omega_o,\mathcal{W}]\}
+\frac{32}{9}\matrixbeta\Omega_o^6,\\
&\mathop{\sum_{\ell,m=1}^{2}\sum_{n=0}^{1}}\limits_{(\ell+m+n=3)}
[\matrixbeta\mathcal{O}^{(\ell)},[\matrixbeta\mathcal{O}^{(m)},h^{(n)}]]\\
&\qquad=-\frac{7}{3}\matrixbeta\{\mathcal{D},[\Omega_o,\mathcal{W}]\}
+\frac{18}{3}\{\mathcal{D},\Omega_o\mathcal{D}\Omega_o\}.
\end{split}
\end{equation}

We note that the operator $\mathcal{D}$ is proportional to
$\matrixboldalpha\cdot\mathbf{E}$, which is of the 1st order of the
electric field as well as the operator $\mathcal{W}$. We find that the
terms $\{\mathcal{D},[\Omega_o,\mathcal{W}]\}$ and
$\{\mathcal{D},\Omega_o\mathcal{D}\Omega_o\}$ in Eq.~(\ref{HFW_5_1}) are proportional to the product of only electric
field, and thus will be neglected. On the other hand, the magnetic
field in $\Omega_o^2$ is also of the 1st order [see Eq.~(\ref{Omega2})]. If we further neglect those terms proportional to
the products of magnetic field and consider the homogeneous
field, $\Omega_o^6$ becomes
\begin{equation}\label{HFW_5_2}
\begin{split}
\frac{\Omega_o^6}{E_g^5}&=\frac{(\Omega_o^2)^3}{E_g^5}\\  
&\approx\frac{1}{32}mc^2\left(\frac{|\boldsymbol{\Pi}|}{mc}\right)^6
-\frac{3}{16}\left(\frac{|\boldsymbol{\Pi}|}{mc}\right)^4\cdot(\boldsymbol{\mu}_p^{\prime
e}\mathbf{B} +\boldsymbol{\mu}_m^{\prime\tilde{e}}\mathbf{E}).
\end{split}
\end{equation}
Therefore, $H_\mathrm{FW}^{(5)}$ with substitution of Eqs.~(\ref{HFW_5_1}) and (\ref{HFW_5_2}) becomes
\begin{equation}\label{HFW5}
\begin{split}
H_\mathrm{FW}^{(5)}&\approx\left(\frac{32}{144}+\frac{32}{18}\right)\frac{\matrixbeta\Omega_o^6}{E_g^5}=\frac{2\matrixbeta\Omega_o^6}{E_g^5}\\
&=\frac{1}{16}mc^2\left(\frac{|\boldsymbol{\Pi}|}{mc}\right)^6
-\frac{3}{8}\left(\frac{|\boldsymbol{\Pi}|}{mc}\right)^4(\boldsymbol{\mu}_p^{\prime
e}\cdot\mathbf{B} +\boldsymbol{\mu}_m^{\prime
\tilde{e}}\cdot\mathbf{E}).
\end{split}
\end{equation}
The first and second terms of Eq.~(\ref{HFW5}) contribute to the
relativistic mass correction and the Zeeman effect, respectively.
For the 6th-order term $H^{(6)}_\mathrm{FW}$, it can be shown that
the commutators of the form
$[\matrixbeta\mathcal{O}^{\ell},\mathcal{O}^{(m)}]$ are given by
\begin{widetext}
\begin{equation}\label{HFW_6_1}
\begin{split}
[\matrixbeta\mathcal{O}^{(1)},\mathcal{O}^{(4)}]&=\frac{8}{15}(\Omega_o^4\mathcal{W}+\Omega_o^3\mathcal{W}\Omega_o+\Omega_o^2\mathcal{W}\Omega_o^2+\Omega_o\mathcal{W}\Omega_0^3+\mathcal{W}\Omega_o^4),\\
[\matrixbeta\mathcal{O}^{(2)},\mathcal{O}^{(3)}]&=-\frac{2}{9}(-\Omega_o^4\mathcal{W}+\Omega_o^3\mathcal{W}\Omega_o+\Omega_o\mathcal{W}\Omega_o^3-\mathcal{W}\Omega_o^4),\\
\end{split}
\end{equation}
where we also have
$[\matrixbeta\mathcal{O}^{(3)},\mathcal{O}^{(2)}]=[\matrixbeta\mathcal{O}^{(2)},\mathcal{O}^{(3)}]$
and
$[\matrixbeta\mathcal{O}^{(4)},\mathcal{O}^{(1)}]=[\matrixbeta\mathcal{O}^{(1)},\mathcal{O}^{(4)}]$.
On the other hand, the commutators of the form
$[\matrixbeta\mathcal{O}^{(\ell)},[\matrixbeta\mathcal{O}^{(m)},h^{(n)}]]$
in $H^{(6)}_\mathrm{FW}$ are given by
\begin{equation}\label{HFW_6_2}
\begin{split}
&[(\matrixbeta\Omega_o)_{(6)},h^{(0)}]=\Omega_0^4\mathcal{W}-4\Omega_o^3\mathcal{W}\Omega_o+6\Omega_o^2\mathcal{W}\Omega_o^2-4\Omega_o\mathcal{W}\Omega_o^3+\mathcal{W}\Omega_o^4,\\
&[\matrixbeta\mathcal{O}^{(1)},[\matrixbeta\mathcal{O}^{(3)},h^{(0)}]]=\frac{1}{6}[\mathcal{D},[[\Omega_o,\mathcal{W}],\Omega_E]],\\
&[\matrixbeta\mathcal{O}^{(2)},[\matrixbeta\mathcal{O}^{(2)},h^{(0)}]]=\frac{16}{9}(\Omega_0^4\mathcal{W}+2\Omega_o^3\mathcal{W}\Omega_o+3\Omega_o^2\mathcal{W}\Omega_o^2+2\Omega_o\mathcal{W}\Omega_o^3+\mathcal{W}\Omega_o^4),\\
&[\matrixbeta\mathcal{O}^{(1)},[\matrixbeta\mathcal{O}^{(2)},h^{(1)}]]=\frac{8}{3}(\Omega_0^4\mathcal{W}+\Omega_o^3\mathcal{W}\Omega_o+\Omega_o^2\mathcal{W}\Omega_o^2+\Omega_o\mathcal{W}\Omega_o^3+\mathcal{W}\Omega_o^4),\\
&[\matrixbeta\mathcal{O}^{(2)},[\matrixbeta\mathcal{O}^{(1)},h^{(1)}]]=\frac{4}{3}(\Omega_0^4\mathcal{W}+\Omega_o^3\mathcal{W}\Omega_o+2\Omega_o^2\mathcal{W}\Omega_o^2+\Omega_o\mathcal{W}\Omega_o^3+\mathcal{W}\Omega_o^4),\\
&[\matrixbeta\mathcal{O}^{(1)},[\matrixbeta\mathcal{O}^{(1)},h^{(2)}]]=\frac{1}{2}[\mathcal{D},[\mathcal{D},\mathcal{W}]].\\
\end{split}
\end{equation}
\end{widetext}
Because $\mathcal{W}$ is of the 1st order of an electric field as
well as $\mathcal{D}$, we can use Eq.~(\ref{HFW_4_2}) to reduce
these equations into a form with only fields of the 1st order. For
example, the term $\Omega_o^3\mathcal{W}\Omega_o$ becomes
$\Omega_o^3\mathcal{W}\Omega_o=\Omega_o^2(\Omega_o\mathcal{W}\Omega_o)\approx
c^4|\boldsymbol{\Pi}|^2(-|\boldsymbol{\Pi}|^2\mathcal{W})$. On the
other hand, $[\mathcal{D},[\mathcal{D},\mathcal{W}]]$ and
$[\mathcal{D},[[\Omega_o,\mathcal{W}],\Omega_E]]$ are neglected
because they are at least of the 2nd order of fields. In that
sense, by the use of Eqs.~(\ref{HFW_6_1}) and (\ref{HFW_6_2}),
one can obtain
\begin{equation}\label{HFW_6_3}
\begin{split}
&[(\matrixbeta\Omega_o)_{(6)},\Omega_E]\approx 16c^4|\boldsymbol{\Pi}|^4\mathcal{W},\\
&\mathop{\sum_{\ell,m=1}^{4}}\limits_{(\ell+m=5)}[\matrixbeta\mathcal{O}^{(\ell)},\mathcal{O}^{(m)}]\approx\frac{128}{45}c^4|\boldsymbol{\Pi}|^4\mathcal{W},\\
&\mathop{\sum_{\ell,m=1}^{3}\sum_{n=0}^{2}}\limits_{(\ell+m+n=4)}[\matrixbeta\mathcal{O}^{(\ell)},[\matrixbeta\mathcal{O}^{(m)},h^{(n)}]]\approx\frac{64}{9}c^4|\boldsymbol{\Pi}|^4\mathcal{W}.
\end{split}
\end{equation}
Therefore, Eq.~(\ref{HFW_list6}) with substitution of
Eq.~(\ref{HFW_6_3}) becomes
\begin{equation}\label{HFW6}
\begin{split}
H_\mathrm{FW}^{(6)}&\approx\frac{1}{E_g^6}\left[\frac{16}{720}+\frac{1}{2}\left(\frac{128}{45}\right)+\frac{1}{2}\left(\frac{64}{9}\right)\right]|\boldsymbol{\Pi}|^4\mathcal{W}\\
&=\frac{5c^4}{E_g^6}|\boldsymbol{\Pi}|^4\mathcal{W}\\
&=\frac{5}{8}\left(\frac{|\boldsymbol{\Pi}|}{mc}\right)^4\left(\frac{\mathcal{W}}{2E_g^2}\right).
\end{split}
\end{equation}
Eq.~(\ref{HFW6}) is the relativistic correction to the spin-orbit
interaction in $H_\mathrm{FW}^{(2)}$. Therefore, $H_\mathrm{FW}^{(1)}$ and
$H_\mathrm{FW}^{(3)}$ and $H_\mathrm{FW}^{(5)}$ are composed of kinetic energy,
interaction energy of Zeeman effect and their relativistic corrections. On the
other hand, $H_\mathrm{FW}^{(2)}$ and $H_\mathrm{FW}^{(4)}$ and $H_\mathrm{FW}^{(6)}$
contain only spin-orbit interaction and its relativistic
corrections.

To simplify the expression of $H_\mathrm{FW}^{(\ell)}$,
we can define a scaled kinetic momentum operator $\boldxi$ as\footnote{It must be stressed that the operator $\boldxi$ does not directly
correspond to the Lorentz boost velocity $\boldsymbol{\beta}$
given in the previous section. The appropriate
transformation between $\boldxi$ and operator for the boost velocity is
considered in Sec.\ref{sec:relation to TBMT}.}
\begin{equation}\label{scaledKM}
\boldxi=\frac{\boldsymbol{\Pi}}{mc}.
\end{equation}
By replacing
$|\boldsymbol{\Pi}|/mc$ with Eq.~(\ref{scaledKM}), Eq.~(\ref{HFW}) with substitution of Eqs.~(\ref{HFW1}), (\ref{HFW2}), (\ref{HFW3}), (\ref{HFW4}), (\ref{HFW5}) and (\ref{HFW6}) becomes a sum of two terms:
\begin{equation}\label{HFW_dipole2}
H_\mathrm{FW}\approx H_\mathrm{orbit}+H_\mathrm{spin},
\end{equation}
where $H_\mathrm{orbit}$ is the kinetic energy plus the potential energy, namely
\begin{equation}\label{HFW_kinetic}
H_\mathrm{orbit}=\matrixbeta
mc^2\left(1+\frac{1}{2}|\boldxi|^2-\frac{1}{8}|\boldxi|^4+\frac{1}{16}|\boldxi|^6\right)+V,
\end{equation}
and $H_\mathrm{spin}$ is the energy of intrinsic dipole moments placing
in electromagnetic fields, namely,
\begin{widetext}
\begin{equation}\label{HFW_dipole3}
\begin{split}
H_\mathrm{spin}&=-\mathbf{E}\cdot
\left[\matrixbeta\boldsymbol{\mu}_p^{\prime\tilde{e}}
+\frac{1}{2}\left(\boldxi\times\boldsymbol{\mu}_m^{\prime e}\right)\right]
-\mathbf{B}\cdot\left[\matrixbeta\boldsymbol{\mu}_m^{\prime e}
-\frac{1}{2}\left(\boldxi\times\boldsymbol{\mu}_p^{\prime\tilde{e}}\right)\right]\\
&+\matrixbeta\left(\frac{1}{2}|\boldxi|^2-\frac{3}{8}|\boldxi|^4\right)
\left(\boldsymbol{\mu}_m^{\prime e}\cdot\mathbf{B}
+\boldsymbol{\mu}_p^{\prime\tilde{e}}\cdot\mathbf{E}\right)
+\left(-\frac{3}{4}|\boldxi|^2+\frac{5}{8}|\boldxi|^4\right)
\left\{-\frac{1}{2}\mathbf{E}\cdot\left(\boldxi\times\boldsymbol{\mu}_m^{\prime e}\right)
-\frac{1}{2}\mathbf{B}\cdot\left(-\boldxi\times\boldsymbol{\mu}_p^{\prime\tilde{e}}\right)
\right\}.
\end{split}
\end{equation}
\end{widetext}

In Sec.~\ref{sec:relation to TBMT}, we will focus on the dipole Hamiltonian
[Eq.~(\ref{HFW_dipole3})]. We will show that Eq.~(\ref{HFW_dipole3}) is in agreement with TBMT equation, provided
that the proper transformation of $\boldxi$ and Lorentz boost
velocity $\boldsymbol{\beta}$ is taken care of.

\subsection{In relation to TBMT equation}\label{sec:relation to TBMT}
We will show that the FW transformation of the Dirac Hamiltonian of a dyon is equivalent to the Hamiltonian obtained from TBMT equation with $g_e=g_{\tilde{e}}=2$. That is, Eq.~(\ref{HFW_kinetic}) is equivalent to Eq.~(\ref{H orbit dyon}) and Eq.~(\ref{HFW_dipole3}) to Eq.~(\ref{H dipole dyon}) with Eq.~(\ref{ThomasF}) and Eq.~(\ref{ThomasF dual}) for $g_e=g_{\tilde{e}}=2$.
However, we must first find the boost velocity in order to compare them.
It must be emphasized that
$\boldsymbol{\beta}$ in TBMT equation is the boost velocity but $\boldxi$ in $H_\mathrm{FW}$ is not. One has to define the boost operator $\operatorboost$ via
\begin{equation}\label{transform}
\boldxi=\frac{\operatorboost}{\sqrt{1-|\operatorboost|^2}},
\qquad
\operatorgamma\equiv\frac{1}{\sqrt{1-|\operatorboost|^2}},
\end{equation}
because the kinetic momentum $\boldsymbol{\Pi}\equiv mc \boldxi=m\mathbf{U}$ and the 4-velocity $U^\alpha=(\gamma c,\gamma\boldsymbol{\beta})$. By using Eq.~(\ref{transform}), the kinetic energy operator Eq.~(\ref{HFW_kinetic}) behaves like
$mc^2\left(1+\frac{1}{2}|\boldxi|^2-\frac{1}{8}|\boldxi|^4+\frac{1}{16}|\boldxi|^4\right)=mc^2(1+\frac{1}{2}|\operatorboost|^2+\frac{3}{8}|\operatorboost|^4+\frac{5}{16}|\operatorboost|^6+o(8))$.
On the other hand, the expansion of Lorentz factor
$\gamma={1}/{\sqrt{1-\beta^2}}$ with respect to small boost
$\beta$ is
$\gamma=1+\frac{1}{2}\beta^2+\frac{3}{8}\beta^4+\frac{5}{16}\beta^6+o(8)$.
This implies that the the kinetic energy operator corresponds to
the classical relativistic energy $\gamma mc^2$, as expected. The
boost operator $\operatorboost$ plays an important role on showing the
equivalence between $H_\mathrm{spin}$ and TBMT Hamiltonian. For an
electron, Eq.~(\ref{HFW_dipole3}) without a magnetic charge
($\tilde{e}=0$) becomes
\begin{widetext}
\begin{equation}\label{HD_electron}
\begin{split}
H_\mathrm{spin}^{(\tilde{e}=0)}&=-\mathbf{E}\cdot
\left[\frac{1}{2}\left(1-\frac{3}{4}|\boldxi|^2+\frac{5}{8}|\boldxi|^4\right)
\left(\boldxi\times\boldsymbol{\mu}_m^{\prime e}\right)\right]-\mathbf{B}\cdot
\left[\matrixbeta\left(1-\frac{1}{2}|\boldxi|^2+\frac{3}{8}|\boldxi|^4\right)
\boldsymbol{\mu}_m^{\prime e}\right]\\
&=-\boldsymbol{\mu}_m^{\prime e}\cdot\left[\matrixbeta\left(1-\frac{1}{2}|\boldxi|^2
+\frac{3}{8}|\boldxi|^4\right)\mathbf{B}-\frac{1}{2}\left(1-\frac{3}{4}|\boldxi|^2
+\frac{5}{8}|\boldxi|^4\right)\boldxi\times\mathbf{E}\right].
\end{split}
\end{equation}
\end{widetext}

The first term in the right hand side of the first equality of
Eq.~(\ref{HD_electron}) is an effective electric dipole moment
caused by the boosted intrinsic spin magnetic moment, which is the
spin-orbit interaction. The second one is the Zeeman term.
Nevertheless, Eq.~(\ref{HD_electron}) provides the relativistic
correction to the Zeeman and spin-orbit interactions. For the
Zeeman term, the non-relativistic limit up to $1/mc$ is
\begin{equation}
H_\mathrm{Zeeman}=-\matrixbeta\boldsymbol{\mu}_m^{\prime e}\cdot\mathbf{B},
\end{equation}
which is the same as the interaction of a classical magnetic
moment and a magnetic field. To the 4th order of $\boldxi$, the
relativistic correction to $H_\mathrm{Zeeman}$ is
\begin{equation}
H_\mathrm{Zeeman}=-\matrixbeta\left(1-\frac{1}{2}|\boldxi|^2+\frac{3}{8}|\boldxi|^4\right)
\boldsymbol{\mu}_m^{\prime e}\cdot\mathbf{B}.
\end{equation}
On the other hand, the spin-orbit interaction denoted as $H_\mathrm{so}$
is
\begin{equation}
\begin{split}
H_\mathrm{so}&=\frac{1}{2}\boldsymbol{\mu}_m^{\prime e}\cdot\boldxi\times\mathbf{E}\\
&=\frac{|e|\hbar}{4m^2c^2}\boldsymbol{\Sigma}\cdot\mathbf{E}\times\boldPi,
\end{split}
\end{equation}
where $e=-|e|$ is used in the second equality, and the
relativistic correction to this term is
\begin{equation}
H_\mathrm{so}=\frac{1}{2}\left(1-\frac{3}{4}|\boldxi|^2+\frac{5}{8}|\boldxi|^4\right)
\boldsymbol{\mu}_m^{\prime e}\cdot\boldxi\times\mathbf{E}.
\end{equation}

We now go back to the discussion of $H_\mathrm{spin}^{\tilde{e}=0}$ and
TBMT equation. In order to compare Eq.~(\ref{HD_electron}) with
TBMT Hamiltonian, we have to transform $\boldxi$ in Eq.~(\ref{HD_electron}) to $\operatorboost$. Using Eq.~(\ref{transform}),
we have
\begin{subequations}
\begin{align}
&\left(1-\frac{1}{2}|\boldxi|^2+\frac{3}{8}|\boldxi|^4\right)=1-\frac{|\operatorboost|^2}{2}-\frac{|\operatorboost|^4}{8}+o(6),\label{expansion2}\\
&\frac{1}{2}\left(1-\frac{3}{4}|\boldxi|^2+\frac{5}{8}|\boldxi|^4\right)|\boldxi|=\frac{|\operatorboost|}{2}-\frac{|\operatorboost|^3}{8}-\frac{|\operatorboost|^5}{16}+o(7).\label{expansion1}
\end{align}
\end{subequations}
The effective spin magnetic moment in TBMT Hamiltonian [Eq.~(\ref{H dipole})] transforms like
$(1/\gamma)\boldsymbol{\mu}_m^{\prime e}$, and we have
\begin{equation}
\frac{1}{\gamma}=1-\frac{\beta^2}{2}-\frac{\beta^4}{8}+o(6),
\end{equation}
which is exactly the same as  Eq.~(\ref{expansion2}) up to terms
of the 4th order in $\beta$. On the other hand, the effective
electric dipole moment transforms like
$(g_e/2-\gamma/(\gamma+1))$, and $g_e$ factor in the Dirac Hamiltonian
is always 2. We obtain
\begin{equation}
\left(1-\frac{\gamma}{1+\gamma}\right)\beta=\frac{\beta}{2}-\frac{\beta^3}{8}-\frac{\beta^5}{16}+o(7),\\
\end{equation}
which is exactly the same as Eq.~(\ref{expansion1}) up to terms of
the 5th order in $\beta$. Because $g_e$ factor equals
2, the longitudinal term $\boldsymbol{\mu}_m^{\prime e}\cdot\boldxi$
disappears in both TBMT Hamiltonian and
$H_\mathrm{spin}^{(\tilde{e}=0)}$. In the following, we will use the
following two approximations directly:
\begin{equation}\label{transgamma}
\begin{split}
&\left(1-\frac{1}{2}|\boldxi|^2+\frac{3}{8}|\boldxi|^4\right)\approx
\frac{1}{\operatorgamma}\,,\\
&\frac{1}{2}\left(1-\frac{3}{4}|\boldxi|^2+\frac{5}{8}|\boldxi|^4\right)|\boldxi|\approx\left(1-\frac{\operatorgamma}{\operatorgamma+1}\right).
\end{split}
\end{equation}
Therefore, we show that up to the fifth order of boost velocity
$\beta$, the Dirac Hamiltonian of an electron is equivalent to the
TBMT Hamiltonian which is obtained from the requirement of
covariance form of classical spin. This implies that in the FW
representation, after summing over all infinite expansion terms,
the Dirac Hamiltonian of an electron would be of the form
\begin{equation}\label{HD_electron2}
H_\mathrm{spin}^{\tilde{e}=0}=-\boldsymbol{\mu}_m^{\prime
e}\cdot\left[\matrixbeta\frac{1}{\operatorgamma}\mathbf{B}-\left(1-\frac{\operatorgamma}{1+\operatorgamma}\right)\operatorboost\times\mathbf{E}\right]
\end{equation}
for the spin part, and of the form
\begin{equation}
H_\mathrm{orbit}=\operatorgamma\matrixbeta mc^2+V
\end{equation}
for orbital part. The effective magnetic field in Eq.~(\ref{HD_electron2}) is the same as Eq.~(\ref{H dipole}) with Eq.~(\ref{ThomasF}) for $g_e=2$.

Furthermore, for the FW transformation of the Dirac Hamiltonian [Eq.~(\ref{HFW_dipole2})], the TBMT
equation can be generalized to include an effective spin magnetic
moment resulting from the boosted intrinsic electric dipole
moment. To the 1st order in $|\boldxi|=|\boldsymbol{\Pi}|/mc$,
we find that the effective dipole moments transform like
\begin{equation}\label{effdipoles}
\begin{split}
(\boldsymbol{\mu}_p^{\tilde{e}})_\mathrm{eff}&\approx
\matrixbeta\boldsymbol{\mu}_p^{\prime\tilde{e}}+\frac{1}{2}
\left(\boldxi\times\boldsymbol{\mu}_m^{\prime e}\right),\\
(\boldsymbol{\mu}_m^e)_\mathrm{eff}&\approx\matrixbeta\boldsymbol{\mu}_m^{\prime e}
-\frac{1}{2}\left(\boldxi\times\boldsymbol{\mu}_p^{\prime\tilde{e}}\right).
\end{split}
\end{equation}
This means that an intrinsic electric dipole moment can result in
an effective magnetic dipole moment when it is moving.
Nevertheless, a moving spin magnetic moment can also intrinsically
induce an effective electric dipole moment. Consider higher orders
of the boost velocity, we rewrite Eq.~(\ref{HFW_dipole3}) as
\begin{widetext}
\begin{equation}
\begin{split}
H_\mathrm{spin}&=-\mathbf{E}\cdot
\left[\matrixbeta\left(1-\frac{1}{2}|\boldxi|^2
+\frac{3}{8}|\boldxi|^4\right)\boldsymbol{\mu}_p^{\prime\tilde{e}}+\frac{1}{2}
\left(1-\frac{3}{4}|\boldxi|^2+\frac{5}{8}|\boldxi|^4\right)
\left(\boldxi\times\boldsymbol{\mu}_m^{\prime e}\right)\right]\\
&~~~~-\mathbf{B}\cdot\left[\matrixbeta
\left(1-\frac{1}{2}|\boldxi|^2+\frac{3}{8}|\boldxi|^4\right)
\boldsymbol{\mu}_m^{\prime e}-\frac{1}{2}\left(1-\frac{3}{4}|\boldxi|^2
+\frac{5}{8}|\boldxi|^4\right)
\left(\boldxi\times\boldsymbol{\mu}_p^{\prime\tilde{e}}\right)\right].
\end{split}
\end{equation}
\end{widetext}
It is shown that intrinsic dipole moments transform like
$\left(1-\frac{1}{2}|\boldxi|^2+\frac{3}{8}|\boldxi|^4\right)\approx
1/\operatorgamma$ and the boosted dipole moments transform as
$\frac{1}{2}\left(1-\frac{3}{4}|\boldxi|^2+\frac{5}{8}|\boldxi|^4\right)\approx(1-\frac{\operatorgamma}{\operatorgamma+1})$
(see Eq.~(\ref{transgamma})). Therefore,
$(\boldsymbol{\mu}_p^{\prime\tilde{e}})_\mathrm{eff}$ and
$(\boldsymbol{\mu}_m^{\prime e})_\mathrm{eff}$ do not form a second rank tensor
in the sense that their transformation is not a covariant form
like Eq.~(\ref{LTdipole}), but the following form:
\begin{equation}
\begin{split}
&(\boldsymbol{\mu}_p^{\tilde{e}})_\mathrm{eff}
\approx\matrixbeta\frac{1}{\operatorgamma}\boldsymbol{\mu}_p^{\prime\tilde{e}}
+\left(1-\frac{\operatorgamma}{\operatorgamma+1}\right)
\left(\operatorboost\times\boldsymbol{\mu}_m^{\prime e}\right),\\
&(\boldsymbol{\mu}_m^e)_\mathrm{eff}
\approx\matrixbeta\frac{1}{\operatorgamma}\boldsymbol{\mu}_m^{\prime e}
-\left(1-\frac{\operatorgamma}{\operatorgamma+1}\right)
\left(\operatorboost\times\boldsymbol{\mu}_p^{\prime\tilde{e}}\right).\\
\end{split}
\end{equation}
This implies that an energy caused by dipole moments in the
description of Dirac Hamiltonian is not simply the contraction of
tensorial dipole density and field tensor:
$H_\mathrm{spin}\neq-\boldsymbol{\mu}_p\cdot\mathbf{E}-\boldsymbol{\mu}_m\cdot\mathbf{B}$,
in which $\boldsymbol{\mu}_p$ and $\boldsymbol{\mu}_m$ transform
as in Eq.~(\ref{LTdipole}). As a result, $H_\mathrm{spin}$ is not
a Lorentz scalar.

In short, in this section we have shown that up to terms of
the 7th order in $1/E_g$, the FW transformation of the Dirac
Hamiltonian of an electron is in agreement with TBMT Hamiltonian [Eq.~(\ref{H dipole})] with $g_e=2$. The result can be generalized to a
particle with an intrinsic electric dipole moment. Because of the
duality of electromagnetic fields, a Dirac dyon would
manifest this feature. Furthermore, we also find the relativistic
corrections to the Zeeman term and spin-orbit interaction

\section{Foldy-Wouthuysen transformation for the Dirac-Pauli Hamiltonian}
\label{sec:Dirac Pauli} In Sec.~\ref{sec:Dirac}, we have shown
that, up to the 7th order in $1/E_g$, the FW transformation of the
Dirac Hamiltonian for a dyon is in agreement with Eq.~(\ref{H
orbit}) and Eq.~(\ref{Thomas}) for $g_e=g_{\tilde{e}}=2$. Since
the Dirac Hamiltonian automatically yields $g_e=2$ and
$g_{\tilde{e}}=2$, the second term in Eq.~(\ref{ThomasF}) and
Eq.~(\ref{ThomasF dual}) vanishes and thus the longitudinal
polarization does not change. In order to see that the
relativistic quantum theory of a spin-$1/2$ particle is in accord
with the TBMT equation even when the change rate of the
longitudinal polarization is concerned, we have to study the
spin-$1/2$ particle with anomalous magnetic dipole moment (AMM)
and anomalous electric dipole moment (AEM).

The relativistic quantum theory of a spin-$/2$ dyon with the
inclusion of AMM and AEM can be described by the Dirac-Pauli
equation \cite{Silenko08, Pauli41}
\begin{equation}
i\hbar\frac{\partial}{\partial t}|\psi\rangle=\mathcal{H}|\psi\rangle,
\end{equation}
where the Dirac-Pauli Hamiltonian $\mathcal{H}$ is the Dirac Hamiltonian $H$ [given in Eq.~(\ref{H})] augmented with the corrections for the AMM and AEM:
\begin{equation}\label{H_ADM}
\mathcal{H}=H+\mu'(-\matrixbeta\boldsymbol{\Sigma}\cdot\mathbf{B}
+i\matrixboldgamma\cdot\mathbf{E})+d'(\matrixbeta\boldsymbol{\Sigma}\cdot\mathbf{E}
+i\matrixboldgamma\cdot\mathbf{B}).
\end{equation}
The coefficients $\mu'$ and $d'$ are defined as follows
\begin{equation}\label{mu-and-d}
\mu'=\left(\frac{g_e}{2}-1\right)\frac{e\hbar}{2mc}, \quad
d'=\left(\frac{g_{\tilde{e}}}{2}-1\right)\frac{\tilde{e}\hbar}{2mc},
\end{equation}
which measures the AMM and AEM, respectively (note $\mu'=0$ for $g_e=2$ and $d'=0$ for $g_{\tilde{e}}=2$).
The $4\times4$ matrices $\matrixbeta\boldsymbol{\Sigma}$ and $\matrixboldgamma$
are defined as
\begin{equation}\label{matrices beta and gamma}
\matrixbeta\boldsymbol{\Sigma}=\left(\begin{array}{cc}
\boldsymbol{\sigma}&0\\
0&-\boldsymbol{\sigma}
\end{array}\right),
\qquad
\matrixboldgamma=\left(\begin{array}{cc}
0&\boldsymbol{\sigma}\\
-\boldsymbol{\sigma}&0
\end{array}\right),
\end{equation}
where $\boldsymbol{\sigma}=(\sigma_x,\sigma_y,\sigma_z)$ are Pauli
matrices. We will see that the Dirac-Pauli Hamiltonian given in
Eq.~(\ref{H_ADM}) is compatible to generic values of $g_e$ and
$g_{\tilde{e}}$ and thus can accommodate AMM and AEM.

In order to obtain the FW transformation of Eq.~(\ref{H_ADM}), we
have to rewrite Eq.~(\ref{H_ADM}) in terms of odd and even
matrices. According to Eq.~(\ref{OE}), we have
\begin{equation}\label{AH}
\mathcal{H}=\matrixbeta mc^2+\Omega_E^A+\Omega_o^A,
\end{equation}
where the superscript $A$ indicates the inclusion of anomalous
dipole moments. We note that $\mu'$ and $d'$ are of the 1st order of
$1/E_g$. Therefore, $\Omega_E^A$ and $\Omega_o^A$ can be written
as
\begin{equation}
\begin{split}
&\Omega_E^A=\Omega_E+\frac{\Omega_E^f}{E_g},\\
&\Omega_o^A=\Omega_o+\frac{\Omega_o^f}{E_g},
\end{split}
\end{equation}
where $\Omega_E$ and $\Omega_o$ are given in Eq.~(\ref{OE1}) and
\begin{equation}
\begin{split}
&\Omega_E^f=\matrixbeta\boldsymbol{\Sigma}\cdot(-\mu''\mathbf{B}+d''\mathbf{E}),\\
&\Omega_o^f=i\matrixboldgamma\cdot(\mu''\mathbf{E}+d''\mathbf{B}),\\
&\mu''=E_g\mu',
\quad
d''=E_gd'.
\end{split}
\end{equation}
The superscript $f$ indicates that these terms are of the 1st order of
electromagnetic fields. Because we consider only those terms
proportional to the 1st order of fields, the products of fields will
be neglected. The validity of Eq.~(\ref{Appendix_mainresult}) is
still true provided that the odd term of the second FW
transformation denoted as $\mathcal{O}'$ in Eq.~(\ref{AH}) starts
from $1/E_g^3$. This can be seen as follows. After the first FW
transformation, $S_1=\matrixbeta\Omega_o^A/E_g$, Eq.~(\ref{AH})
becomes
\begin{equation}
\mathcal{H}'=\frac{\matrixbeta}{2}E_g+h+\mathcal{O},
\end{equation}
where $h$ and $\mathcal{O}$ are even and odd terms, respectively.
It can be shown that the odd term $\mathcal{O}$ can be written as
\begin{equation}\label{AH2}
\mathcal{O}=\frac{\mathcal{O}^{(1)}}{E_g}+\frac{\mathcal{O}^{(2)}}{E^2_g}+\frac{\mathcal{O}^{(3)}}{E^3_g}+\frac{\mathcal{O}^{(4)}}{E^4_g}+o(\frac{1}{E_g^5}),
\end{equation}
where the corresponding $\mathcal{O}^{(n)},~n=1,2,3,4$, are given by
\begin{equation}\label{AO}
\begin{split}
&\mathcal{O}^{(1)}=\matrixbeta[\Omega_o,\Omega_E],\\
&\mathcal{O}^{(2)}=2c\matrixbeta\matrixeta(-\mu''\mathbf{B}+d''\mathbf{E})\cdot\boldPi-\frac{4}{3}\Omega_o^3,\\
&\mathcal{O}^{(3)}=-\frac{4}{3}\left(\Omega_o\{\Omega_o,\Omega_o^f\}+\Omega_o^f\Omega_o^2\right)+\frac{1}{6}\matrixbeta[\Omega_o,\mathcal{W}],\\
&\mathcal{O}^{(4)}=-\frac{4}{3}c^3\matrixbeta\matrixeta(-\mu''\mathbf{B}+d''\mathbf{E})\cdot\boldPi|\boldPi|^2+\frac{8}{15}\Omega_o^5,
\end{split}
\end{equation}
where the matrix $\matrixeta$ is defined as
\begin{equation}\label{matrix eta}
\matrixeta=\left(\begin{array}{cc}
0&-\mathbf{1}\\
\mathbf{1}&0
\end{array}\right).
\end{equation}
If $g_e$ and $g_{\tilde e}$ are both equal to 2, then Eq.~(\ref{AO}) goes back to
Eq.~(\ref{listO}). Using the second FW transformation
$S_2=\matrixbeta\mathcal{O}/E_g$ on Eq.~(\ref{AH2}), we can obtain
the other odd term denoted as $\mathcal{O}'$. The 1st order term
$\mathcal{O}'^{(1)}$ is zero because $\mathcal{O}$ starts form
the 1st order of $1/E_g$ at least. $\mathcal{O}'^{(2)}$ can be
written as
$\mathcal{O}'^{(2)}=[\matrixbeta\mathcal{O}^{(1)},h^{(0)}]$.
Nevertheless, we have
$\mathcal{O}^{(1)}=\matrixbeta[\Omega_o,\Omega_E]=\matrixbeta
i\hbar\matrixboldalpha\cdot(e\mathbf{E}+\tilde{e}\mathbf{B})$ and
$h^{(0)}=\Omega_E=V$ is a scalar that commutes with
$\mathcal{O}^{(1)}$, and thus $\mathcal{O}'^{(2)}$ vanishes. This
implies that Eq.~(\ref{Appendix_mainresult}) is still valid in
this case. The matrix $h$ in Eq.~(\ref{AH2}) is given by
\begin{equation}\label{AMMAEM_h}
\begin{split}
h=&\Omega_E^A+\frac{1}{2E_g}[\matrixbeta\Omega_o^A,\Omega_o^A]+\frac{1}{2E_g^2}[(\matrixbeta\Omega_o^A)_{(2)},\Omega_E^A]\\
&+\frac{1}{8E_g^3}[(\matrixbeta\Omega_o^A)_{(3)},\Omega_o^A]+\frac{1}{24}[(\matrixbeta\Omega_o^A)_{(4)},\Omega_E^A]\\
&+\frac{1}{144E_g^5}[(\matrixbeta\Omega_o^A)_{(5)},\Omega_o^A]+\frac{1}{720E_g^6}[(\matrixbeta\Omega_o^A)_{(6)},\Omega_E^A].
\end{split}
\end{equation}
To obtain $h'$, we need extra corrections to $h$, as shown in Eq.~(\ref{Appendix_h'expand}). The resulting Hamiltonian can be
written as
\begin{equation}
\mathcal{H}_\mathrm{FW}=H_\mathrm{FW}+\mathcal{H}_\mathrm{FW1}+\mathcal{H}_\mathrm{FW2}.
\end{equation}
The Hamiltonian $H_\mathrm{FW}$ is given in
Eq.~(\ref{HFW_dipole2}), $\mathcal{H}_\mathrm{FW1}$ contains those
terms proportional to $(-\mu''\mathbf{B}+d''\mathbf{E})$, and
$\mathcal{H}_\mathrm{FW2}$ contains those terms proportional to
$(\mu''\mathbf{E}+d''\mathbf{B})$. Focusing on the term
proportional to $(-\mu'\mathbf{B}+d'\mathbf{E})$, namely,
$[(\matrixbeta\Omega_o^A)_{(n=2,4)},\Omega_E^A]$ in Eq.
(\ref{AMMAEM_h}). For $n=2$, we have
\begin{equation}
\begin{split}
[(\matrixbeta\Omega_o^A)_{(2)},\Omega_E^A]&=[\matrixbeta\Omega_o^A,[\matrixbeta\Omega_o^A,\Omega_E^A]]\\
&=[[\Omega_o^A,\Omega_E^A],\Omega_o^A]\\
&=\mathcal{W}+\frac{\mathcal{W}^f}{E_g}\,,
\end{split}
\end{equation}
where $\mathcal{W}$ is given in Eq.~(\ref{DandW}) and
$\mathcal{W}^f$ is defined as
\begin{equation}\label{Wf}
\begin{split}
\mathcal{W}^f&=[[\Omega_o,\Omega_E^f],\Omega_o]\\
&=-4c^2\matrixbeta\boldsymbol{\Sigma}\cdot\boldPi(-\mu''\mathbf{B}+d''\mathbf{E})\cdot\boldPi,
\end{split}
\end{equation}
which is proportional to electric and magnetic fields. By using
Eq.~(\ref{Wf}), the term
$[(\matrixbeta\Omega_o^A)_{(4)},\Omega_E^A]$ can be written as
\begin{equation}\label{Wf2}
\begin{split}
&[(\matrixbeta\Omega_o^A)_{(4)},\Omega_E^A]\\
&=[\matrixbeta\Omega_o^A,[\matrixbeta\Omega_o^A,\mathcal{W}+\frac{\mathcal{W}^f}{E_g}]]\\
&=[[\Omega_o,\mathcal{W}],\Omega_o]-\frac{4c^2}{E_g}\mathcal{W}^f|\boldPi|^2.
\end{split}
\end{equation}
The first term of Eq.~(\ref{Wf2}) is just one of the 4th order
terms of $H_\mathrm{FW}^{(4)}$ shown in Sec.~\ref{sec:dipoles}. It
is important to note that the second term of Eq.~(\ref{Wf2}) is
collected in $h^{(5)}$, not $h^{(4)}$. It can be shown that the
only term that contributes to
$\boldsymbol{\Sigma}\cdot\boldPi(-\mu''\mathbf{B}+d''\mathbf{E})\cdot\boldPi|\boldPi|^2$
is $[\matrixbeta\mathcal{O}^{(2)},\mathcal{O}^{(2)}]$ that is the
correction term in $h'^{(5)}$ (see Eq.~(\ref{Appendix_h'expand})).
Using $\mathcal{O}^{(2)}$ in Eq.~(\ref{AO}), we have
\begin{equation}
\begin{split}
&\frac{1}{2}[\matrixbeta\mathcal{O}^{(2)},\mathcal{O}^{(2)}]\\
&=\matrixbeta(\mathcal{O}^{(2)})^2\\
&=\matrixbeta\left(2c\matrixbeta\matrixeta(-\mu''\mathbf{B}+d''\mathbf{E})\cdot\boldPi-\frac{4}{3}\Omega_o^3\right)^2\\
&=\matrixbeta\frac{16}{9}\Omega_o^6-\frac{8}{3}c^4(-\mu''\mathbf{B}+d''\mathbf{E})\cdot\boldPi|\boldPi|^2[\matrixeta,\matrixalpha_{\ell}]\Pi_{\ell}.
\end{split}
\end{equation}
It can be shown that
$[\matrixeta,\matrixalpha_{\ell}]=-2\matrixbeta\Sigma_{\ell}$, and thus we
obtain
\begin{eqnarray}\label{Wf3}
&&\frac{1}{2}[\matrixbeta\mathcal{O}^{(2)},\mathcal{O}^{(2)}]\\
&=&\matrixbeta\frac{16}{9}\Omega_o^6+\frac{16}{3}c^4\matrixbeta\boldsymbol{\Sigma}\cdot\boldPi(-\mu''\mathbf{B}+d''\mathbf{E})\cdot\boldPi|\boldPi|^2.\nonumber
\end{eqnarray}
Consider those terms proportional to
$(-\mu''\mathbf{B}+d''\mathbf{E})$: one comes form Eq.~(\ref{Wf})
in the corresponding term and the other is obtained form
$\Omega_E^f$, Eq.~(\ref{Wf2}) and Eq.~(\ref{Wf3}). Using the
definitions of $\boldxi=\boldPi/mc$ (see Eq.~(\ref{transform})),
$\mu''=E_g\mu'$ and $d''=E_gd'$, one obtain
\begin{equation}\label{H_FW1_1}
\begin{split}
\mathcal{H}_\mathrm{FW1}&=\frac{1}{2E_g^3}\mathcal{W}^f+\matrixbeta\boldsymbol{\Sigma}\cdot(-\mu'\mathbf{B}+d'\mathbf{E})\\
&~~~~+\frac{c^4}{E_g^5}\left(\frac{16}{3}+\frac{16}{24}\right)\matrixbeta\boldsymbol{\Sigma}\cdot\boldPi(-\mu''\mathbf{B}+d''\mathbf{E})\cdot\boldPi|\boldPi|^2\\
&=\left(-\frac{1}{2}+\frac{3}{8}|\boldxi|^2\right)\matrixbeta\boldsymbol{\Sigma}\cdot\boldxi(-\mu'\mathbf{B}+d'\mathbf{E})\cdot\boldxi\\
&~~~~+\matrixbeta\boldsymbol{\Sigma}\cdot(-\mu'\mathbf{B}+d'\mathbf{E}).
\end{split}
\end{equation}

For $g_e=2$ and $g_{\tilde{e}}=2$,
$\mathcal{H}_\mathrm{FW1}$ vanishes. We now transform $\boldxi$ in
Eq.~(\ref{H_FW1_1}) in terms of the boost velocity $\operatorboost$. By
using the transformation between $\boldxi$ and $\operatorboost$ [see
Eq.~(\ref{transform})], it can be shown that
$\left(\frac{1}{2}-\frac{3}{8}|\boldxi|^2\right)|\boldxi|^2\approx\frac{\operatorgamma}{1+\operatorgamma}|\operatorboost|^2$.
Substituting AMM coefficient
$\mu'=\left(\frac{g_e}{2}-1\right)\frac{e\hbar}{2mc}$ and AEM
coefficient
$d'=\left(\frac{g_{\tilde{e}}}{2}-1\right)\frac{\tilde{e}\hbar}{2mc}$
into Eq.~(\ref{H_FW1_1}), we find that $\mathcal{H}_\mathrm{FW1}$
can be written as
\begin{widetext}
\begin{equation}
\begin{split}
\mathcal{H}_\mathrm{FW1}&
=-\frac{\operatorgamma}{\operatorgamma+1}\matrixbeta\boldsymbol{\Sigma}
\cdot\operatorboost(-\mu'\mathbf{B}+d'\mathbf{E})\cdot\operatorboost
+\matrixbeta\boldsymbol{\Sigma}\cdot(-\mu'\mathbf{B}+d'\mathbf{E})\\
&=-\frac{\operatorgamma}{\operatorgamma+1}\matrixbeta
\left[-\left(\frac{g_e}{2}-1\right)\boldsymbol{\mu}^{\prime e}_m\cdot
\operatorboost\mathbf{B}\cdot\operatorboost-\left(\frac{g_{\prime\tilde{e}}}{2}
-1\right)\boldsymbol{\mu}^{\prime\tilde{e}}_p\cdot\operatorboost\mathbf{E}
\cdot\operatorboost\right]+\matrixbeta\left[-\left(\frac{g_e}{2}-1\right)
\boldsymbol{\mu}_m^{\prime e}\cdot\mathbf{B}-\left(\frac{g_{\tilde{e}}}{2}-1\right)
\boldsymbol{\mu}_p^{\prime \tilde{e}}\cdot\mathbf{E}\right].
\end{split}
\end{equation}
\end{widetext}
On the other hand, the terms proportional to
$(\mu''\mathbf{E}+d''\mathbf{B})$ correspond to
$\{\Omega_o,\Omega_o^f\}=2c\matrixbeta\boldsymbol{\Sigma}\cdot\boldPi\times(\mu''\mathbf{E}+d''\mathbf{B})$.
Similar to the derivation for $\mathcal{H}_\mathrm{FW1}$, we
collect all terms proportional to $\{\Omega_o,\Omega_o^f\}$ and
obtain
\begin{equation}
\mathcal{H}_\mathrm{FW2}=\matrixbeta\{\Omega_o,\Omega_o^f\}\frac{1}{E_g^2}\left(1-\frac{1}{2}|\boldxi|^2+\frac{3}{8}|\boldxi|^4\right).
\end{equation}
Using $\mu''=E_g\mu'$ and $d''=E_gd'$, we have
\begin{equation}
\mathcal{H}_\mathrm{FW2}=\left(1-\frac{1}{2}|\boldxi|^2+\frac{3}{8}|\boldxi|^4\right)\boldsymbol{\Sigma}\cdot\boldxi\times(\mu'\mathbf{E}+d'\mathbf{B}).
\end{equation}
By using Eq.~(\ref{transform}), we find that
\begin{eqnarray}
\mathcal{H}_\mathrm{FW2}&=&
\boldsymbol{\Sigma}\cdot\operatorboost\times(\mu'\mathbf{E}+d'\mathbf{B})\\
&=&\left(\frac{g_e}{2}-1\right)\boldsymbol{\mu}^{\prime e}_m
\cdot(\operatorboost\times\mathbf{E})
-\left(\frac{g_{\tilde{e}}}{2}-1\right)\boldsymbol{\mu}^{\prime\tilde{e}}_p\cdot(\operatorboost\times\mathbf{B}).\nonumber
\end{eqnarray}
To focus on the interaction of dipole moments and external fields,
we have to combine $H_\mathrm{spin}$, $\mathcal{H}_\mathrm{FW1}$
and $\mathcal{H}_\mathrm{FW2}$ together. After a straightforward
calculations, we find that
\begin{widetext}
\begin{equation}\label{mainresult}
\begin{split}
&H_\mathrm{spin}+\mathcal{H}_\mathrm{FW1}+\mathcal{H}_\mathrm{FW2}\\
&=-\boldsymbol{\mu}_m^{\prime e}\cdot\left\{\left(\frac{g_e}{2}-1+\frac{1}{\operatorgamma}\right)\matrixbeta\mathbf{B}-\left(\frac{g_e}{2}-\frac{\operatorgamma}{\operatorgamma+1}\right)\operatorboost\times\mathbf{E}-\frac{\operatorgamma}{\operatorgamma+1}\left(\frac{g_e}{2}-1\right)\operatorboost(\mathbf{B}\cdot\operatorboost)\right\}\\
&~~~-\boldsymbol{\mu}_p^{\prime\tilde{e}}\cdot\left\{\left(\frac{g_{\tilde{e}}}{2}-1+\frac{1}{\operatorgamma}\right)\matrixbeta\mathbf{E}+\left(\frac{g_{\tilde{e}}}{2}-\frac{\operatorgamma}{\operatorgamma+1}\right)\operatorboost\times\mathbf{B}-\frac{\operatorgamma}{\operatorgamma+1}\left(\frac{g_{\tilde{e}}}{2}-1\right)\operatorboost(\mathbf{E}\cdot\operatorboost)\right\}.\\
\end{split}
\end{equation}
\end{widetext}
Eq.~(\ref{mainresult}) is in agreement with Eq.~(\ref{H dipole
dyon}) when the replacement $\gamma\rightarrow\operatorgamma$ and
the duality transformation for electromagnetic fields are used.
Without magnetic charge, Eq.~(\ref{mainresult}) coincides with
Eq.~(\ref{ThomasF}) for arbitrary values of $g_e$. The
dual part of the TBMT equation for spin is also obtained.

In short, the whole derivations in this section have assumed that
electromagnetic fields are static and homogeneous. Therefore, we
show that up to terms of the 7th order in $1/E_g$, the FW
transformation including anomalous dipole moments coincides with
the TBMT equation for the spin-$1/2$ particle with arbitrary $g_e$
and $g_{\tilde e}$.

\section{Conclusions and discussion}\label{sec:conclusions}
To investigate the low-energy limit of the relativistic quantum
theory for a spin-$1/2$ charged particle, which is described by
the Dirac equation, we perform a series of successive FW
transformations on the Dirac Hamiltonian up to terms of the 7th
order in $1/E_g$. Assuming the electromagnetic fields are static
and homogeneous, and taking care of the relation between the
kinematic momentum $\boldsymbol{\Pi}$ used in the Dirac
Hamiltonian and the boost velocity $\boldsymbol{\beta}$ used in
the TBMT equation, we show that the resulting FW transformation of
the Dirac Hamiltonian is in agreement with the classical orbital
Hamiltonian $H_\mathrm{orbit}$ plus the TBMT Hamiltonian
$H_\mathrm{spin}$ with the gyromagnetic ratio $g_e$ being equal to
2. Through electromagnetic duality, this can be generalized for a
spin-$1/2$ dyon, which has both electric and magnetic charges and
thus possesses both intrinsic magnetic dipole moment
$\boldsymbol{\mu}_m^{\prime e}$ and intrinsic electric dipole
moment $\boldsymbol{\mu}_p^{\prime\tilde{e}}$.

To affirm the consistency between the low-energy limit of the
relativistic quantum theory and the classical counterpart to a
broader extent, we consider the relativistic quantum theory for a
spin-$1/2$ dyon with arbitrary values of the gyromagnetic and
gyroelectric ratios, which is described by the Dirac-Pauli
equation, namely, the Dirac equation with augmentation for AMM and
AEM. Up the 7th order in $1/E_g$ again, we show that the FW
transformation of the Dirac-Pauli Hamiltonian is also in accord
with $H_\mathrm{orbit}+H_\mathrm{spin}$.

Many phenomena regarding spin dynamics have been observed and can
be explained by the TBMT equation. These include the anomalous
Zeeman effect, spin-orbit interaction, Thomas precession and
change rate of the longitudinal polarization (see Sec.~11.11 of
\cite{Jackson} for a brief review). The TBMT equation is however
derived merely by the requirement of covariance without invoking
any first principles. By studying the FW transformation of the
Dirac Hamiltonian and the Dirac-Pauli Hamiltonian, we have shown
that the TBMT equation as a phenomenological formula is in fact
supported by the first principle of the fundamental relativistic
quantum theory as a low-energy limit. (The relativistic quantum
theory further requires the spin to be quantized as
$\mathbf{s}=\hbar\,\boldsymbol{\sigma}/2$; this result cannot be
obtained at the phenomenological level.) Therefore, the
correspondence principle is again shown to be established.

By far, the agreement between the Dirac/Dirac-Pauli equation and
the orbital equation plus the TBMT equation is only proven up to
the 7th order in $1/E_g$. Further research is needed to
investigate the FW transformation to higher orders and a generic
expression for the FW transformation at an arbitrary order could
be obtained by mathematical induction. If this is the case,
performing successive FW transformations ad infinitum is expected
to yield the result in precise agreement with the orbital equation
plus the TBMT equation.

Furthermore, the assumption of static and homogeneous fields can
be released. In time-varying and/or inhomogeneous fields, the TBMT
equation has to be generalized to allow gradient force terms like
$(\boldsymbol{\mu}_m\cdot\boldsymbol{\nabla})\mathbf{B}$ and the
FW transformation of the Dirac/Dirac-Pauli Hamiltonian shall yield
the corresponding terms accordingly. The gradient force terms
should not be missing, as
$(\boldsymbol{\mu}_m\cdot\boldsymbol{\nabla})\mathbf{B}$ is used
in the Stern-Gerlach experiment to separate spin-up and spin-down
particles. Furthermore, the detailed investigation for the dipole
moments interacting with the time-variation of electromagnetic
fields may predict new physics. However, the calculation for the
FW transformation will be much more complicated if the fields are
non-static and inhomogeneous.

\section*{ACKNOWLEDGMENTS}
The authors would like to thank Chih-Wei Chang for valuable
discussions and D.W.C. is grateful to Jiun-Huei Wu for the warm
hospitality during his visit at National Taiwan University. T.W.C.
is supported by the financial support from the National Science
Council and NCTS of Taiwan; D.W.C. is supported by the NSFC Grant
No. 10675019 and the financial support of Grants No. 20080440017
and No. 200902062 from the China Postdoctoral Science Foundation.

\appendix

\section{Foldy-Wouthuysen transformation}\label{sec:FW transform}
In this appendix, we expand the Dirac Hamiltonian up to terms
of the 7th order in $1/E_g$ with $E_g=2mc^2$. The Dirac Hamiltonian
can be separated in to two parts. One is the even operator denoted
as $\Omega_E$, which commutes with $\matrixbeta$, and the other is
odd operator $\Omega_o$, which anti-commutes with $\matrixbeta$:
\begin{equation}
\begin{split}
[\matrixbeta,\Omega_E]=0,\\
\{\matrixbeta,\Omega_o\}=0.
\end{split}
\end{equation}
The Dirac Hamiltonian can be written as
\begin{equation}
H=\frac{\matrixbeta}{2}E_g+\Omega_E+\Omega_o,
\end{equation}
where $\Omega_E=e\phi+\tilde{e}\tilde{\phi}$ and
$\Omega_o=c\matrixboldalpha\cdot\boldPi$. The kinetic momentum $\boldPi$
is
$\boldPi=\mathbf{p}-\frac{e}{c}\mathbf{A}-\frac{\tilde{e}}{c}\widetilde{\mathbf{A}}$.
The first transformation operator can be written as
\begin{equation}
U_{1}=e^{S_1}, \quad S_1=\matrixbeta\Omega_o/E_g.
\end{equation}
The Dirac Hamiltonian under the unitary transformation $U_1$ can be
written as
\begin{widetext}
\begin{equation}\label{Appendix_HFW1}
H_{1}=U_1HU^{-1}_1=\frac{\beta}{2}E_g+\Omega_E+\sum_{n=1}^{\infty}\frac{1}{E_g^n}\left\{\left(\frac{1}{n!}-\frac{1}{(n+1)!}\right)[(\matrixbeta\Omega_o)_{(n)},\Omega_o]+\frac{1}{n!}[(\matrixbeta\Omega_o)_{(n)},\Omega_E]\right\},
\end{equation}
\end{widetext}
where the subscript $n$ at $(\matrixbeta\Omega)_{(n)}$ is defined
as, for example,
$[(\matrixbeta\Omega_o)_{(3)},\Omega_o]=[\matrixbeta\Omega_o,[\matrixbeta\Omega_o,[\matrixbeta\Omega_o,\Omega_E]]]$.
Equation~(\ref{Appendix_HFW1}) can be again separated into odd and
even parts. The even part of Eq.~(\ref{Appendix_HFW1}) denoted as
$h$ can be written as
\begin{equation}
h=h^{(0)}+\sum_{n=1}^{\infty}\frac{h^{(n)}}{E_g^n},
\end{equation}
where $h^{(0)}=\Omega_E$ and
\begin{equation}\label{Appendix_h}
\begin{split}
&h^{(n=1,3,5,\cdots)}=\left(\frac{1}{n!}-\frac{1}{(n+1)!}\right)[(\matrixbeta\Omega_o)_{(n)},\Omega_o],\\
&h^{(n=2,4,6,\cdots)}=\frac{1}{n!}[(\matrixbeta\Omega_o)_{(n)},\Omega_E].
\end{split}
\end{equation}
The odd part of Eq.~(\ref{Appendix_HFW1}) denoted as $\mathcal{O}$
can be written as
\begin{equation}
\mathcal{O}=\sum_{n=1}^{\infty}\frac{\mathcal{O}^{(n)}}{E_g^n},
\end{equation}
where
\begin{equation}\label{Appendix_unprimeO}
\begin{split}
&\mathcal{O}^{(n=1,3,5,\cdots)}=\frac{1}{n!}[(\matrixbeta\Omega_o)_{(n)},\Omega_E],\\
&\mathcal{O}^{(n=2,4,6,\cdots)}=\left(\frac{1}{n!}-\frac{1}{(n+1)!}\right)[(\matrixbeta\Omega_o)_{(n)},\Omega_o].
\end{split}
\end{equation}
Therefore, Eq.~(\ref{Appendix_HFW1}) becomes
\begin{equation}
H_1=\frac{\matrixbeta}{2}E_g+h+\mathcal{O},
\end{equation}
where $h$ contains those terms with only even matrices and
$\mathcal{O}$ contains only odd matrices. The second
transformation denoted as $U_2=\exp(S_2)$, where $S_2$ is
$S_2=\matrixbeta\mathcal{O}/E_g$. We have
\begin{widetext}
\begin{equation}\label{Appendix_HFW2}
\begin{split}
H_2=U_2H_1U^{-1}_2&=\frac{\beta}{2}E_g+h+\sum_{n=1}^{\infty}\frac{1}{E_g^n}\left\{\left(\frac{1}{n!}-\frac{1}{(n+1)!}\right)[(\matrixbeta\mathcal{O})_{(n)},\mathcal{O}]+\frac{1}{n!}[(\matrixbeta\mathcal{O})_{(n)},h]\right\}\\
&=\frac{\beta}{2}E_g+h'+\mathcal{O}',
\end{split}
\end{equation}
where $h'$ and $\mathcal{O}'$ are the new even and odd parts, respectively, of
the right hand side of the first equality. The even part
of Eq.~(\ref{Appendix_HFW2}) denoted as $h'$ can be written as
\begin{equation}\label{Appendix_h'}
\begin{split}
h'&=h+\left\{\begin{array}{c}
\displaystyle \sum_{n=1,3,5\cdots}\frac{1}{E_g^n}\left(\frac{1}{n!}-\frac{1}{(n+1)!}\right)[(\matrixbeta\mathcal{O})_{(n)},\mathcal{O}]\\
\\
\displaystyle
\sum_{n=2,4,6\cdots}\frac{1}{E_g^n}\frac{1}{n!}[(\matrixbeta\mathcal{O})_{(n)},h]
\end{array}\right.\\
&=h'^{(0)}+\sum_{m=1}^{\infty}\frac{h'^{(m)}}{E_g^{m}}.
\end{split}
\end{equation}
The odd term $\mathcal{O}'$ in Eq.~(\ref{Appendix_HFW2}) is given
by
\begin{equation}\label{Appendix_primeO}
\begin{split}
\mathcal{O}'&=\left\{\begin{array}{c}
\displaystyle \sum_{n=2,4,6\cdots}\frac{1}{E^n_g}\left(\frac{1}{n!}-\frac{1}{(n+1)!}\right)[(\matrixbeta\mathcal{O})_{(n)},\mathcal{O}]\\
\\
\displaystyle
\sum_{n=1,3,5\cdots}\frac{1}{E_g^n}\frac{1}{n!}[(\matrixbeta\mathcal{O})_{(n)},h]
\end{array}\right.\\
&=\sum_{m=3}^{\infty}\frac{\mathcal{O}'^{(m)}}{E_g^m}.
\end{split}
\end{equation}
Firstly, we note that $\mathcal{O}'^{(1)}$ and
$\mathcal{O}'^{(2)}$ in Eq.~(\ref{Appendix_primeO}) vanish as we
have $m$ starting from $3$. For the former result, the reason is
that the lowest order of the unprimed odd term $\mathcal{O}$ is 1,
and thus in the second line of the first equality in
Eq.~(\ref{Appendix_primeO}), the primed odd term $\mathcal{O}'$ is
at least of the 2nd order. On the other hand, the explicit form of
$\mathcal{O}'^{(2)}$ can be written as
$\mathcal{O}'^{(2)}=[\matrixbeta\mathcal{O}^{(1)},h^{(0)}]$.
However, $\mathcal{O}^{(1)}$ is given by
Eq.~(\ref{Appendix_unprimeO}) for $n=1$, namely,
$\mathcal{O}^{(1)}=\matrixbeta[\Omega_o,\Omega_E]
=ic\hbar\matrixboldalpha\cdot(e\mathbf{E}+\tilde{e}\widetilde{\mathbf{E}})$,
which commutes with
$h^{(0)}=\Omega_E=e\phi+\tilde{e}\tilde{\phi}$. If we perform the
third transformation which is $S_3=\matrixbeta\mathcal{O}'/E_g$ at
$H_2=\frac{\matrixbeta E_g}{2}+h'+\mathcal{O}'$, we will obtain a
new even term $h''$ as well as the new odd term $\mathcal{O}''$:
$H_3=U_{3}H_2U^{-1}_3=\frac{\matrixbeta}{2}E_g+h''+\mathcal{O}''$.
The even term $h''$ is given by
\begin{equation}\label{Appendix_h''}
\begin{split}
h''&=h'+\left\{\begin{array}{c}
\displaystyle \sum_{n=1,3,5\cdots}\frac{1}{E_g^n}\left(\frac{1}{n!}-\frac{1}{(n+1)!}\right)[(\matrixbeta\mathcal{O}')_{(n)},\mathcal{O}']\\
\\
\displaystyle
\sum_{n=2,4,6\cdots}\frac{1}{E_g^n}\frac{1}{n!}[(\matrixbeta\mathcal{O}')_{(n)},h']
\end{array}\right.\\
&=h''^{(0)}+\sum_{m=1}^{\infty}\frac{h''^{(m)}}{E_g^{m}}.
\end{split}
\end{equation}
The odd term $\mathcal{O}''$ is given by
\begin{equation}\label{Appendix_prime2O}
\begin{split}
\mathcal{O}''&=\left\{\begin{array}{c}
\displaystyle \sum_{n=2,4,6\cdots}\frac{1}{E^n_g}\left(\frac{1}{n!}-\frac{1}{(n+1)!}\right)[(\matrixbeta\mathcal{O}')_{(n)},\mathcal{O}']\\
\\
\displaystyle
\sum_{n=1,3,5\cdots}\frac{1}{E_g^n}\frac{1}{n!}[(\matrixbeta\mathcal{O}')_{(n)},h']
\end{array}\right.\\
&=\sum_{m=4}^{\infty}\frac{\mathcal{O}''^{(m)}}{E_g^m}.
\end{split}
\end{equation}
The odd term $\mathcal{O}''$ starts form $m=4$.  Obviously,
$\mathcal{O}^{(1)}$ is zero because $\mathcal{O}'^{(m)}$ starts
from $m=3$. This can also be seen as follows. The explicit form of
the next three terms of $\mathcal{O}''^{(m)}$ are
\begin{equation}
\begin{split}
&\mathcal{O}''^{(2)}=[\matrixbeta\mathcal{O}'^{(1)},h'^{(0)}],\\
&\mathcal{O}''^{(3)}=[\matrixbeta\mathcal{O}'^{(2)},h'^{(0)}]+[\matrixbeta\mathcal{O}'^{(1)},h'^{(1)}],\\
&\mathcal{O}''^{(4)}=[\matrixbeta\mathcal{O}'^{(1)},h'^{(2)}]+[\matrixbeta\mathcal{O}'^{(2)},h'^{(1)}]+[\matrixbeta\mathcal{O}'^{(3)},h'^{(0)}].\\
\end{split}
\end{equation}
Because $\mathcal{O}'^{(1)}$ and $\mathcal{O}'^{(2)}$ are zero,
$\mathcal{O}''^{(2)}$ and $\mathcal{O}''^{(3)}$ vanish. It can be
shown that $\mathcal{O}''^{(4)}$ does not vanish. Using
Eq.~(\ref{Appendix_h''}), $h''^{(m)}$ for $m=1,2,\cdots 6$ are
given by
\begin{equation}
\begin{split}
&h''^{(0)}=h'^{(0)},\\
&h''^{(1)}=h'^{(1)},\\
&h''^{(2)}=h'^{(2)},\\
&h''^{(3)}=h'^{(3)}+\left(1-\frac{1}{2!}\right)[\matrixbeta\mathcal{O}'^{(1)},\mathcal{O}'^{(1)}],\\
&h''^{(4)}=h'^{(4)}+\left(1-\frac{1}{2!}\right)\left([\matrixbeta\mathcal{O}'^{(1)},\mathcal{O}'^{(2)}]+[\matrixbeta\mathcal{O}'^{(2)},\mathcal{O}'^{(1)}]\right)+\frac{1}{2!}[\matrixbeta\mathcal{O}'^{(1)},[\matrixbeta\mathcal{O}'^{(1)},h'^{(0)}]],\\
&h''^{(5)}=h'^{(5)}+\left(1-\frac{1}{2!}\right)\mathop{\sum_{\ell,m=1}^{3}}\limits_{(\ell+m=4)}[\matrixbeta\mathcal{O}'^{(\ell)},\mathcal{O}'^{(m)}]+\mathop{\frac{1}{2!}\sum_{\ell,m=1}^2\sum_{n=0}^1}\limits_{(\ell+m+n=3)}[\matrixbeta\mathcal{O}'^{(\ell)},[\matrixbeta\mathcal{O}'^{(m)},h'^{(n)}]],\\
&h''^{(6)}=h'^{(6)}+\left(1-\frac{1}{2!}\right)\mathop{\sum_{\ell,m=1}^{4}}\limits_{(\ell+m=5)}[\matrixbeta\mathcal{O}'^{(\ell)},\mathcal{O}'^{(m)}]+\mathop{\frac{1}{2!}\sum_{\ell,m=1}^3\sum_{n=0}^2}\limits_{(\ell+m+n=4)}[\matrixbeta\mathcal{O}'^{(\ell)},[\matrixbeta\mathcal{O}'^{(m)},h'^{(n)}]].\\
\end{split}
\end{equation}
Because $\mathcal{O}'^{(1)}$ and $\mathcal{O}'^{(2)}$ are zero, we
have $h''^{(3)}=h'^{(3)}$ and $h''^{(4)}=h'^{(4)}$, since the
constraints $\ell+m=4$ and $\ell+m=5$ imply
$(\ell,m)=\{(1,3),(2,2),(3,1)\}$ and
$(\ell,m)=\{(1,4),(2,3),(3,2),(4,1)\}$, respectively. The
commutator $[\matrixbeta\mathcal{O}'^{(\ell)},\mathcal{O}'^{(m)}]$
in $h''^{(5)}$ and $h''^{(6)}$ vanishes. Consider the term
$[\matrixbeta\mathcal{O}'^{(\ell)},[\matrixbeta\mathcal{O}'^{(m)},h'^{(n)}]]$
in $h''^{(5)}$ and $h''^{(6)}$ subject to the constraints
$\ell+m+n=3$ and $\ell+m+n=4$, respectively. For $n=0$, we have
$(\ell,m)=\{(1,2),(2,1)\}$ and $(\ell,m)=\{(1,3),(2,2),(3,1)\}$
and thus
$[\matrixbeta\mathcal{O}'^{(\ell)},[\matrixbeta\mathcal{O}'^{(m)},h'^{(n=0)}]]$
vanishes in $h''^{(5)}$ and $h''^{(6)}$. For $n=1$ and $n=2$, the
term
$[\matrixbeta\mathcal{O}'^{(\ell)},[\matrixbeta\mathcal{O}'^{(m)},h'^{(n)}]]$
still vanishes.

Therefore, up to terms of the 7th order in $\frac{1}{E_g}$, we
obtain an important result
\begin{equation}\label{Appendix_mainresult}
h''^{(n)}=h'^{(n)},
\quad
n=1,2,\cdots 6,
\end{equation}
and $h'^{(n)}$ (i.e., Eq.~(\ref{Appendix_h'})) is given by
\begin{equation}\label{Appendix_h'expand}
\begin{split}
&h'^{(0)}=h^{(0)},\\
&h'^{(1)}=h^{(1)},\\
&h'^{(2)}=h^{(2)},\\
&h'^{(3)}=h^{(3)}+\left(1-\frac{1}{2!}\right)[\matrixbeta\mathcal{O}^{(1)},\mathcal{O}^{(1)}],\\
&h'^{(4)}=h^{(4)}+\left(1-\frac{1}{2!}\right)\left([\matrixbeta\mathcal{O}^{(1)},\mathcal{O}^{(2)}]+[\matrixbeta\mathcal{O}^{(2)},\mathcal{O}^{(1)}]\right)+\frac{1}{2!}[\matrixbeta\mathcal{O}^{(1)},[\matrixbeta\mathcal{O}^{(1)},h^{(0)}]],\\
&h'^{(5)}=h^{(5)}+\left(1-\frac{1}{2!}\right)\mathop{\sum_{\ell,m=1}^{3}}\limits_{(\ell+m=4)}[\matrixbeta\mathcal{O}^{(\ell)},\mathcal{O}^{(m)}]+\mathop{\frac{1}{2!}\sum_{\ell,m=1}^2\sum_{n=0}^1}\limits_{(\ell+m+n=3)}[\matrixbeta\mathcal{O}^{(\ell)},[\matrixbeta\mathcal{O}^{(m)},h^{(n)}]],\\
&h'^{(6)}=h^{(6)}+\left(1-\frac{1}{2!}\right)\mathop{\sum_{\ell,m=1}^{4}}\limits_{(\ell+m=5)}[\matrixbeta\mathcal{O}^{(\ell)},\mathcal{O}^{(m)}]+\mathop{\frac{1}{2!}\sum_{\ell,m=1}^3\sum_{n=0}^2}\limits_{(\ell+m+n=4)}[\matrixbeta\mathcal{O}^{(\ell)},[\matrixbeta\mathcal{O}^{(m)},h^{(n)}]].\\
\end{split}
\end{equation}
Therefore, in order to obtain the FW transformation up to
$1/E_g^7$, we need to know four odd terms $\mathcal{O}^{(n)}$ for
$n=1,2,3,4$. On the other hand, if we perform the transformation
again by using $S_4=\matrixbeta\mathcal{O}''/E_g$, since
$\mathcal{O}''^{(m)}$ starts form $m=4$ (i.e.
$\mathcal{O}''^{(m)}=0$ for $m=1,2,3$), the transformation $S_4$
does not change Eq.~(\ref{Appendix_mainresult}). The resulting
$\mathcal{O}'''^{(m)}$ will start from at least $m=5$. To bring
the odd term to the 7th order, we need
$S_5=\matrixbeta\mathcal{O}'''/E_g$ and
$S_6=\matrixbeta\mathcal{O}''''/E_g$. However, the two
transformations also do not change the validity of
Eq.~(\ref{Appendix_mainresult}). Therefore the resulting
Hamiltonian can be written as
\begin{equation}
\begin{split}
H_\mathrm{FW}&=U_\mathrm{FW}HU^{-1}_\mathrm{FW}\\
&=\frac{\matrixbeta}{2}E_g+\sum_{n=0}^{6}H^{(n)}_\mathrm{FW}+o(1/E_g^7).
\end{split}
\end{equation}
By using Eqs.~(\ref{Appendix_h}), (\ref{Appendix_h'expand}) and
(\ref{Appendix_mainresult}), after straightforward calculations it
can be shown that
\begin{subequations}\label{ALLHFW}
\begin{align}
&H^{(1)}_\mathrm{FW}=\frac{\matrixbeta\Omega_o^2}{E_g},\\
&H^{(2)}_\mathrm{FW}=\frac{1}{E_g^2}\left(\frac{\mathcal{W}}{2}\right),\\
&H^{(3)}_\mathrm{FW}=\frac{1}{E_g^3}\left\{-\matrixbeta\Omega_o^4+\matrixbeta\left(\matrixbeta\mathcal{D}\right)^2\right\},\\
&H^{(4)}_\mathrm{FW}=\frac{1}{E_g^4}\left(\frac{1}{24}[[\Omega_o,\mathcal{W}],\Omega_o]-\frac{4}{3}[\mathcal{D},\Omega_o^3]\right),\\
&H^{(5)}_\mathrm{FW}=\frac{1}{E_g^5}\left\{\frac{1}{144}[(\matrixbeta\Omega_o)_{(5)},\Omega_o]+\frac{1}{2}\mathop{\sum_{\ell,m=1}^{3}}\limits_{(\ell+m=4)}[\matrixbeta\mathcal{O}^{(\ell)},\mathcal{O}^{(m)}]+\frac{1}{2}\mathop{\sum_{\ell,m=1}^{2}\sum_{n=0}^{1}}\limits_{(\ell+m+n=3)}[\matrixbeta\mathcal{O}^{(\ell)},[\matrixbeta\mathcal{O}^{(m)},h^{(n)}]]\right\},\\
&H^{(6)}_\mathrm{FW}=\frac{1}{E_g^6}\left\{\frac{1}{720}[(\matrixbeta\Omega_o)_{(6)},\Omega_o]+\frac{1}{2}\mathop{\sum_{\ell,m=1}^{4}}\limits_{(\ell+m=5)}[\matrixbeta\mathcal{O}^{(\ell)},\mathcal{O}^{(m)}]+\frac{1}{2}\mathop{\sum_{\ell,m=1}^{3}\sum_{n=0}^{2}}\limits_{(\ell+m+n=4)}[\matrixbeta\mathcal{O}^{(\ell)},[\matrixbeta\mathcal{O}^{(m)},h^{(n)}]]\right\},\\
\end{align}
\end{subequations}
where $\mathcal{D}\equiv[\Omega_o,\Omega_E]$ and
$\mathcal{W}\equiv[\mathcal{D},\Omega_o]$.
\end{widetext}

\section{Derivation of Equation~(\ref{HFW_4_2(2)})}\label{sec:derivation}
The explicit form of $\Omega_o\mathcal{W}\Omega_o$
[Eq.~(\ref{HFW_4_2})] plays an important role in obtaining the
correct coefficient of each term in $H_\mathrm{FW}^{(6)}$. By
definition, $\Omega_o=c\,\matrixboldalpha\cdot\boldPi$ and
$\mathcal{W}=[[\Omega_o,\Omega_E],\Omega_o]
=[[c\matrixboldalpha\cdot\boldPi,V],c\,\matrixboldalpha\cdot\boldPi]$,
where $V=e\phi+\tilde{e}\widetilde{\phi}$. In the derivation for
Eq.~(\ref{HFW_4_2}), electromagnetic fields are assumed to be
homogeneous and static. This means that the terms involving
gradient of fields are neglected. Therefore, we have
\begin{equation}\label{A2_1}
\frac{1}{c^2}\Omega_o\mathcal{W}\Omega_o=-2c^2\hbar\epsilon_{pqr}\mathcal{E}_q\Pi_i\Pi_r\Pi_j(\matrixalpha_i\Sigma_p\matrixalpha_j),
\end{equation}
where $\mathcal{E}_q\equiv(eE_q+\tilde{e}\tilde{E}_q)$ and
\begin{equation}
\mathcal{W}=-2c^2\hbar\boldsymbol{\Sigma}\cdot(\boldsymbol{\mathcal{E}}\times\boldsymbol{\Pi})
\end{equation}
is used. Furthermore, it can be shown that
\begin{equation}\label{A2_PiPj}
[\Pi_i,\Pi_j]=\frac{i\hbar}{c}\,\epsilon_{ijk}\mathcal{B}_k,
\end{equation}
were $\mathcal{B}_k=eB_k+\tilde{e}\tilde{B}_k$. By using
$[\matrixalpha_i,\sigma_p]=2i\epsilon_{ipm}\matrixalpha_m$ and
$\epsilon_{pqr}\epsilon_{ipm}=\delta_{qm}\delta_{ri}-\delta_{qi}\delta_{rm}$,
Eq.~(\ref{A2_1}) can be written as
\begin{equation}\label{A2_2}
\begin{split}
&\frac{1}{c^2}\Omega_o\mathcal{W}\Omega_o\\
&=-2\hbar\mathcal{E}_q(2i\matrixalpha_q\matrixalpha_j|\boldPi|^2\Pi_j-2i\matrixalpha_r\matrixalpha_j\Pi_q\Pi_r\Pi_j\\
&~~+\epsilon_{pqr}\Sigma_p\Pi_i\Pi_r\Pi_i+i\epsilon_{pqr}\epsilon_{ijm}\Sigma_m\Sigma_p\Pi_i\Pi_r\Pi_j).
\end{split}
\end{equation}
By using
$\matrixalpha_q\matrixalpha_j=\delta_{qj}+i\epsilon_{qj\ell}\Sigma_{\ell}$,
the first term of Eq.~(\ref{A2_2}) becomes
\begin{equation}\label{A2_3}
2i\matrixalpha_q\matrixalpha_j|\boldPi|^2\Pi_j=2i|\boldPi|^2\Pi_q-2\epsilon_{qj\ell}\Sigma_{\ell}|\boldPi|^2\Pi_j.
\end{equation}
On the other hand, the second term of Eq.~(\ref{A2_2}) can be
written as
\begin{equation}\label{A2_4}
-2i\matrixalpha_r\matrixalpha_j\Pi_q\Pi_r\Pi_j=-2i\Pi_q|\boldPi|^2+\frac{2i\hbar}{c}\Sigma_{\ell}\Pi_q\mathcal{B}_{\ell},
\end{equation}
where Eq.~(\ref{A2_PiPj}) is used. The third term of Eq.~(\ref{A2_2}) can be written as
\begin{equation}\label{A2_5}
\begin{split}
&\epsilon_{pqr}\Sigma_p\Pi_i\Pi_r\Pi_i\\
&=\epsilon_{pqr}\Sigma_p\Pi_i\left(\frac{i\hbar}{c}\epsilon_{ri\ell}\mathcal{B}_{\ell}+\Pi_i\Pi_r\right)\\
&=\frac{i\hbar}{c}\left(\Sigma_i\Pi_i\mathcal{B}_q-\Sigma_{\ell}\Pi_{q}\mathcal{B}_{\ell}\right)+\epsilon_{pqr}\Sigma_p|\boldPi|^2\Pi_r.
\end{split}
\end{equation}
The fourth term of Eq.~(\ref{A2_2}) becomes
\begin{equation}\label{A2_6}
\begin{split}
&i\epsilon_{pqr}\epsilon_{ijm}\Sigma_m\Sigma_p\Pi_i\Pi_r\Pi_j\\
&=i\epsilon_{pqr}\epsilon_{ijm}\Sigma_m\Sigma_p\Pi_i\left(\frac{i\hbar}{c}\epsilon_{rj\ell}\mathcal{B}_{\ell}+\Pi_j\Pi_r\right)\\
&=-\frac{\hbar}{c}\epsilon_{pqr}(2B_{\ell}\Sigma_{\ell}\Pi_r-\mathcal{B}_i\Pi_i\Sigma_r+\mathcal{B}_m\Sigma_m)\Sigma_p,
\end{split}
\end{equation}
where $\epsilon_{ijm}\Pi_i\Pi_j=\frac{i\hbar}{c}\mathcal{B}_m$ and
$\epsilon_{ijm}\epsilon_{rj\ell}=\delta_{ir}\delta_{m\ell}-\delta_{i\ell}\delta_{mr}$
are used. We note that there is a field $\mathcal{E}_q$ in the Eq.~(\ref{A2_2}). By neglecting the product of fields
$\mathcal{E}_i\mathcal{B}_j$, Eq.~(\ref{A2_2}) with substitutions
of Eqs.~(\ref{A2_3}), (\ref{A2_4}), (\ref{A2_5}) and (\ref{A2_6})
becomes
\begin{equation}
\begin{split}
\frac{1}{c^2}\Omega_o\mathcal{W}\Omega_o&\approx-2c^2\hbar\mathcal{E}_q(-2\epsilon_{qj\ell}\Sigma_{\ell}|\boldPi|^2\Pi_j+\epsilon_{pqr}\Sigma_p|\boldPi|^2\Pi_r)\\
&=-|\boldPi|^2(-2c^2\epsilon_{pqr}\Sigma_p\mathcal{E}_q\Pi_r)\\
&=-|\boldPi|^2\mathcal{W},
\end{split}
\end{equation}
and we have
\begin{equation}
\Omega_o\mathcal{W}\Omega_o\approx-c^2|\boldPi|^2\mathcal{W}.
\end{equation}

\end{document}